\newfont{\vs}{cmssdc10 scaled 1050}

\documentclass[useAMS,usenatbib,usegraphicx]{mn2e}
\usepackage{xspace,amsmath}
\usepackage{color,epsfig,amssymb,lscape}
\usepackage{util}
\usepackage{array,colortbl}
\usepackage{epstopdf}
\usepackage[dvips]{graphicx}
\usepackage{times,url,graphicx}
\usepackage{longtable}
\usepackage{txfonts}
\usepackage{natbib}
\usepackage[figuresright]{rotating}
\usepackage{subfigure}
\usepackage{multirow}
\usepackage{rotating}
\usepackage{afterpage}

\title[Spatially resolved IFS of the ionized gas in IZw18]{Spatially resolved integral field spectroscopy of the ionized gas in IZw18\thanks{Based on observations collected at the Centro Astron\'omico Hispano Alem\'an (CAHA) at Calar Alto, operated jointly by the Max-Planck-Institut f\"ur Astronomie and the Instituto de Astrof\'isica de Andaluc\'ia (CSIC).}}
\author[C. Kehrig et
al.]{C. Kehrig$^{1}$\thanks{E-mail:kehrig@iaa.es}, J.M. V\'ilchez$^{1}$, E. P\'erez-Montero$^{1}$, J. Iglesias-P\'aramo$^{1,2}$, 
\newauthor
J. D. Hern\'andez-Fern\'andez$^{3}$, S. Duarte Puertas$^{1}$, J. Brinchmann$^{4,5}$, F. Durret$^{6}$, D. Kunth$^{6}$\\
$^{1}$ Instituto de Astrof\'{\i}sica de Andaluc\'{\i}a, CSIC, Apartado de correos 3004, 18080 Granada, Spain \\
$^{2}$ Estaci\'on Experimental de Zonas Aridas (CSIC), Ctra. de Sacramento s/n, La Caada, Almer\'{\i}a, Spain \\
$^{3}$ Departamento de Astronomia, Instituto de Astronomia,Geof\'{\i}sica
e Ci\^encias Atmosf\'ericas da Universidade de S\~ao Paulo, Rua do
Mat\~ao 1226, Cidade \\ Universit\'aria 05508-090 S\~ao Paulo, Brazil \\
$^{4}$ Leiden Observatory, Leiden University, PO Box 9513, 2300 RA Leiden, The Netherlands \\
$^{5}$ Instituto de Astrof\'{\i}sica e Ci{\^e}ncias do Espa\c{c}o,
Universidade do Porto, CAUP, Rua das Estrelas, P-4150-762 Porto, Portugal \\
$^{6}$ Institut d'Astrophysique de Paris, UMR 7095 CNRS, Universit\'e Pierre \& Marie Curie, 98 bis boulevard Arago, 75014 Paris, France}

\begin{document}

\date{Accepted Date. Received Date; in original Date}

\pagerange{\pageref{firstpage}--\pageref{lastpage}} \pubyear{2016}

\maketitle

\label{firstpage}

\begin{abstract}

We present a detailed 2D study of the ionized ISM of IZw18 using new
PMAS-IFU optical observations. IZw18 is a high-ionization galaxy which
is among the most metal-poor starbursts in the local Universe. This
makes IZw18 a local benchmark for understanding the properties most
closely resembling those prevailing at distant starbursts. Our
IFU-aperture ($\sim$ 1.4 $\times$ 1.4 kpc$^{2}$) samples the entire
IZw18 main body and an extended region of its ionized gas. Maps of
relevant emission lines and emission line ratios show that
higher-excitation gas is preferentially located close to the NW knot
and thereabouts. We detect a Wolf-Rayet feature near the NW knot. We
derive spatially resolved and integrated physical-chemical properties
for the ionized gas in IZw18. We find no dependence between the
metallicity-indicator R$_{23}$ and the ionization parameter (as traced
by [OIII]/[OII]) across IZw18. Over $\sim$ 0.30 kpc$^{2}$, using the
[OIII]$\lambda$4363 line, we compute $T_{\rm e}$[OIII] values ($\sim$
15000 - 25000 K), and oxygen abundances are derived from the direct
determinations of $T_{\rm e}$[OIII]. More than 70$\%$ of the
higher-$T_{\rm e}$[OIII] ($\gtrsim$ 22000 K) spaxels are
HeII$\lambda$4686-emitting spaxels too. From a statistical analysis,
we study the presence of variations in the ISM physical-chemical
properties. A galaxy-wide homogeneity, across hundreds of parsecs, is
seen in O/H. Based on spaxel-by-spaxel measurements, the
error-weighted mean of 12 + log(O/H) = 7.11 $\pm$ 0.01 is taken as the
representative O/H for IZw18.  Aperture effects on the derivation of
O/H are discussed. Using our IFU data we obtain, for the first time,
the IZw18 integrated spectrum.
\end{abstract}
%

\section{Introduction}

HII galaxies typically have low masses and blue optical colours, and
are the most metal deficient starbursts (sites with intense
massive star-formation) in the local
Universe \citep[e.g.,][]{HH99,K00,W04,K06,I12}. The optical spectra of HII galaxies
are dominated by strong nebular emission-lines formed via the
ionization of the gas caused by hot massive
stars \citep[e.g.,][]{K04,C09mrk1418,C10,epm10,V14}. The cosmological
relevance of local metal-poor starburst galaxies has been underscored
by the existence of high-redshift starbursts and of the
expected primitive galaxies \citep[e.g.,][]{H98,sch03}. The latter are
supposed to host massive population III stars (PopIII-stars). Such
stars are believed to be the first
generation of stars in the Universe and their energetic UV-light could
have contributed to the reionization of the Universe \citep[e.g.,][]{TS00,BR13}. Investigating low
metallicity HII galaxies in the local Universe can therefore impact
our understanding of distant galaxies and galaxy evolution.

The nearby HII galaxy IZw18 is one of the best analogues of primeval
galaxies accessible to detailed study \citep[e.g.,][]{PO12,LEB13}.
This galaxy keeps attracting attention since its discovery \citep{Z66}
mainly because of its extremely low metallicity \citep[Z $\sim$ 1/40
solar metallicity\footnote{Solar metallicity
Z$_{\odot}$=0.0134 \citep{asplund09}};
e.g.,][]{SS72,KS86,pagel92,jvm98,IT99}. IZw18 also shows very blue
colours and is rather dust
deficient \citep[e.g.,][]{vanzee98,F14}. Optical CCD images of IZw18
have revealed a very complex structure with the presence of shells,
loops and filaments of ionized gas illustrating the effects of
powerful stellar winds in the IZw18's main body and towards its
extended gaseous halo
\citep[e.g.,][]{D89,HT95,cannon02}.  Several studies of the ionized
interstellar medium (ISM) in IZw18 have already been performed, though, despite the complex morphology of the ISM, 
all these works are mostly based on single-aperture/long-slit
spectroscopy of the central star-forming (SF) knots of IZw18. 
\citep[e.g.,][]{LE79,SK93,G97,IT99,TI05,OMB08}. 

The importance of integral
field spectroscopy (IFS), in comparison to single-aperture/long-slit
spectroscopy, for improving our understanding of the warm ISM
conditions in different systems has been demonstrated in the literature 
\citep[e.g.,][]{K08,C09mrk409,J10,AMI11,EPM13,P13}. A detailed bidimensional spectroscopic study of
IZw18, because of its extremely low metal content and high-ionization
gas, is relevant to shed light on the ISM properties from
galaxies in the intermediate/high-z Universe. In \citet[][]{K15}, we derive the total He{\sc ii}-ionizing photon flux in IZw18 and find that peculiar very hot stars, similar to PopIII-stars, are needed to explain the observed nebular He{\sc ii} emission. As far as we know, here we present the first study based on IFS to investigate the
spatially resolved physical-chemical properties (e.g., electron
temperature, gaseous metal abundances, excitation) and the
spatial correlations that may hold for the warm ISM in IZw18.  Our
integral field unit (IFU) data reveal the spatially resolved
ionization structure for the ionized gas of IZw18, which provide useful boundary
conditions for photoionization models at the lowest metallicity regime
\citep[e.g.,][]{M15}. Moreover, taking advantage of the IFU data, we
derive the integrated physical-chemical characteristics for selected regions, including the IZw18 integrated spectrum, that can be helpful
to interpret high-z emission-line galaxies.  We also discuss the
significance of the observed spatial variations of electron
temperatures and the oxygen abundance in terms of the observed
spatially resolved ionization structure. This can impose constraints
on the models for metal dispersal and mixing in HII galaxies, and on
the chemical
evolution models for conditions close to the ones that prevailed in
the primordial Universe \citep[e.g.,][]{R04,Y11,RH13}.

\begin{figure}
\centering
\includegraphics[width=7.5cm]{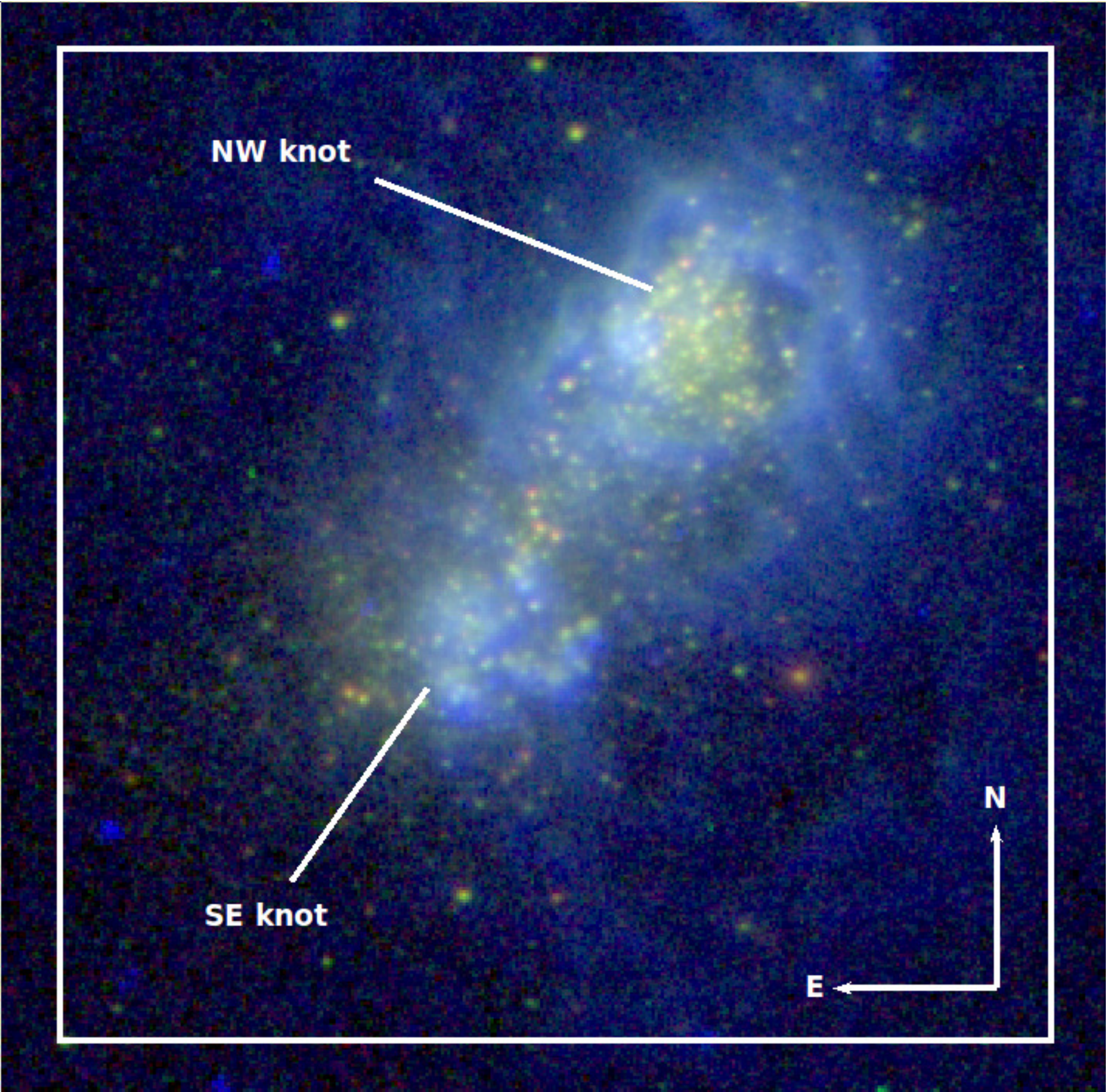}
\caption{Colour composite {\it Hubble Space Telescope} image of IZw18 in three bandpasses (blue
  = WFPC2/F658N, green = WFPC2/F555W, red = WFPC2/F814W, i.e. H$\alpha$-V-I). The observed field of view (FOV) of PMAS (16'' $\times$ 16'') is represented by the white box. North is up and east is to the left.}
\label{pmas_ifu}
\end{figure}

General properties of IZw18 are presented in Table~\ref{sample}.  In
Fig.~\ref{pmas_ifu}, we show a three-colour composite image of IZw18
from the {\it Hubble Space Telescope}/WFPC2. In this image we can see
the two main SF regions of IZw18, usually referred to as the northwest
(NW) and southeast (SE) components; they are separated by an angular
distance of $\sim$ 6 arcsec, and dominate in brightness.

The paper is organized as follows. In Section 2, we report
observations and data reduction. Flux measurements and emission line
intensity maps are presented in Section 3.  In Sections 4 and 5, we
show the 2D view of the ionization structure and nebular properties,
respectively. Section 6 discusses the spatial variation of chemical
abundances and physical conditions over our IFU-aperture. In Section
7, we present the integrated properties from  selected regions of IZw18. Finally,
Section 8 summarizes the main conclusions derived from this work.

\begin{table}
\caption{General Properties of IZw18\label{sample}}
\centering
\begin{tabular}{lc} \hline
Parameter      &IZw18 \\ \hline
Other designation  & Mrk~116, UGCA~166       \\
Morphological type$^{a}$ & i0-type BCD \\
R.A. (J2000.0)  & 09h 34m 02.0s         \\
DEC. (J2000.0)      & +55d 14' 28''        \\
redshift   & 0.0025               \\
D$^{b}$(Mpc) &18.2    \\
Scale (pc/$\prime\prime$)     & 88  \\
(B-R)$^a$    & -0.18  $\pm$0.08         \\
{\it u}$^c$ (mag) & 16.00 $\pm$ 0.05       \\
{\it g}$^c$ (mag) & 15.83 $\pm$ 0.05   \\
{\it r}$^c$ (mag) & 15.87 $\pm$ 0.05    \\
{\it i}$^c$ (mag) & 16.57 $\pm$ 0.05    \\
{\it z}$^c$ (mag) & 16.59 $\pm$ 0.10    \\
A$_{V}$$^d$(mag) & 0.091   \\
\hline 
\end{tabular}
\begin{flushleft}
$^{a}$ From \cite{gildepaz03}; $^{b}$ Distance based on the
tip of the Red Giant Branch distance from \cite{A07}; $^{c}$ From \cite{b14}; $^{d}$ Galactic extinction from \cite{schlegel98} 
\end{flushleft}
\end{table}

\section{Observations and data reduction}

IZw18 was observed in 2012 December, with the IFU Potsdam
Multi-Aperture Spectrophotometer \citep[PMAS;][]{M05,M10}, attached to 
the 3.5 m telescope at the Calar Alto Observatory. A grating with 500
grooves per mm was used during the observing night; this provides a
spectral range from $\sim$ 3640 to 7200 \AA~ with a linear dispersion
$\sim$ 2 \AA/pixel and an effective spectral resolution of $\sim$
3.6 \AA. The observations were performed using the PMAS lens array
mode for which the field-of-view (FOV) is formed by 256 fibres. Each
fibre has a spatial sampling of 1'' $\times$ 1" on the sky resulting
in a FOV of 16'' $\times$ 16'' ($\sim$ 1.4 kpc $\times$ 1.4 kpc at the
distance of 18.2 Mpc) (see Fig.~\ref{pmas_ifu}).

We observed a total of 2.5 hours on the galaxy, with the integration
time split into six exposures of 1500 s each; sky frames were taken
moving the IFU away from the target position
in order to provide the sky background emission to be subtracted from
the target spectra. All science frames were observed at airmasses
$\sim$ 1.1 to minimize the effects due to differential atmospheric
refraction. Additionally, all necessary calibration frames (exposures
of arc lamps and of continuum lamps) were obtained. Observations of
the spectrophotometric standard star Feige 34 were obtained throughout
the observing night to flux calibrate the data.

The data reduction was performed following the procedure described in
\cite{K13}.  We have reduced the IFU data using the P3d \citep{sandin10}
and IRAF\footnote{IRAF is distributed by the National Optical Astronomical Observatories, which are operated by the Association of Universities for Research in Astronomy, Inc., under
cooperative agreement with the National Science Foundation.}
software. We checked the accuracy of the wavelength calibration by measuring the central
wavelength of the [O{\sc i}]$\lambda$5577 \AA~sky line in all fibres and
found a standard deviation of $\sim$ 0.30 \AA.

\section{Flux measurements and emission line intensity maps}\label{flux_measurements}

In this work we measure emission line fluxes with the SPLOT
routine in IRAF, by integrating all the line flux between two points given
by the position of a local continuum. The continuum level is estimated
by visually placing the graphics cursor at both sides of each
line. This process was repeated several times for each emission line by
varying the continuum position. We take the mean and the standard
deviation of the repeated measurements as the final flux of each line
and its associated uncertainty, respectively. The relative errors in
the line intensities are typically $\sim$ 5 per cent for the bright
lines (e.g., H$\beta$; [O{\sc iii}]$\lambda$5007; H$\alpha$). Typical
uncertainties for the faintest lines ([O{\sc iii}]$\lambda$4363;
[N{\sc ii}]$\lambda$6584; [O{\sc i}]$\lambda$6300) are about $\sim$ 15-25 per cent
and may reach up to $\sim$ 30-40 per cent in some fibres.

\begin{figure*}
\center
\includegraphics[bb=1 1 505 405,width=0.45\textwidth,clip]{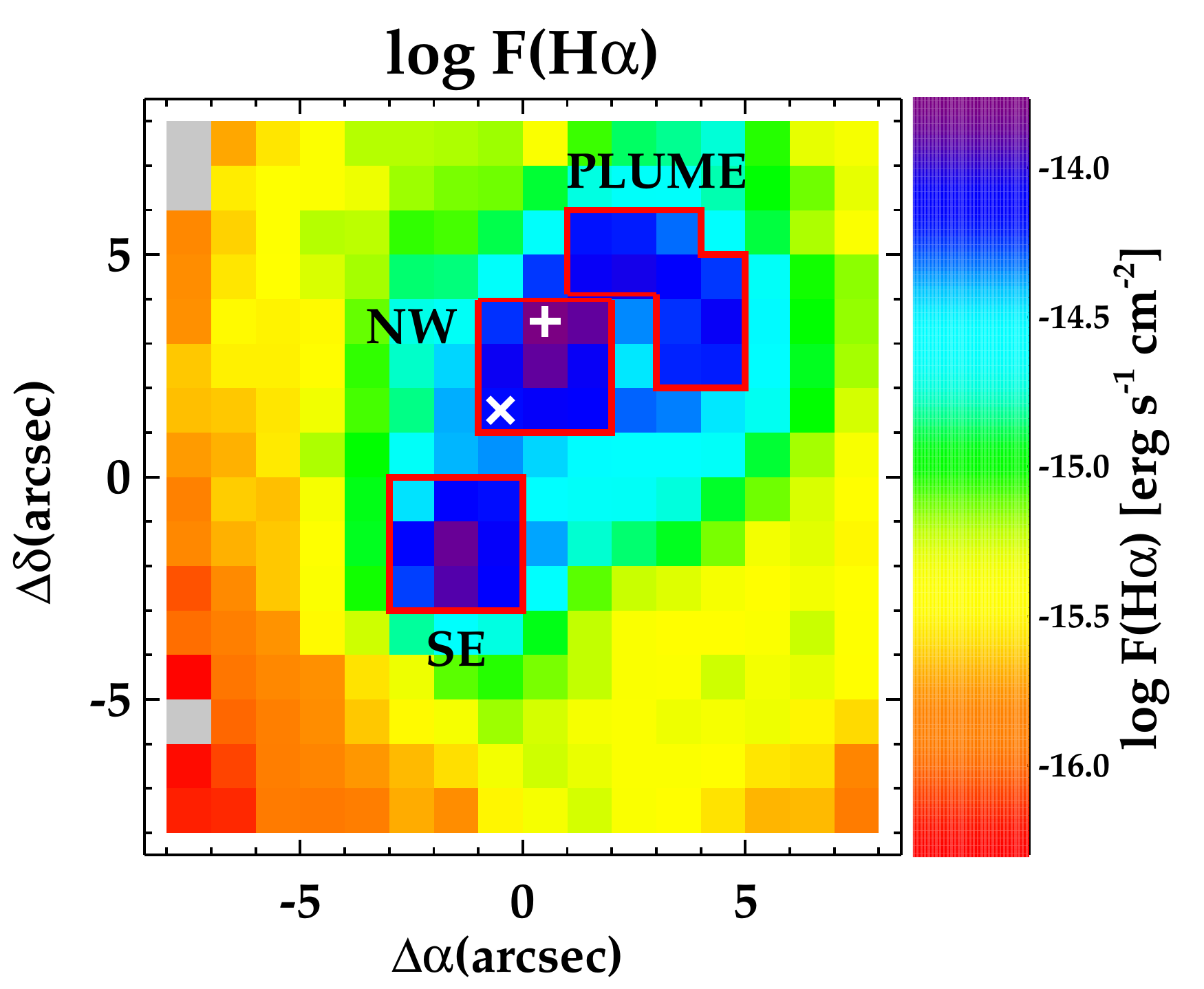}
\includegraphics[bb=1 1 505 405,width=0.45\textwidth,clip]{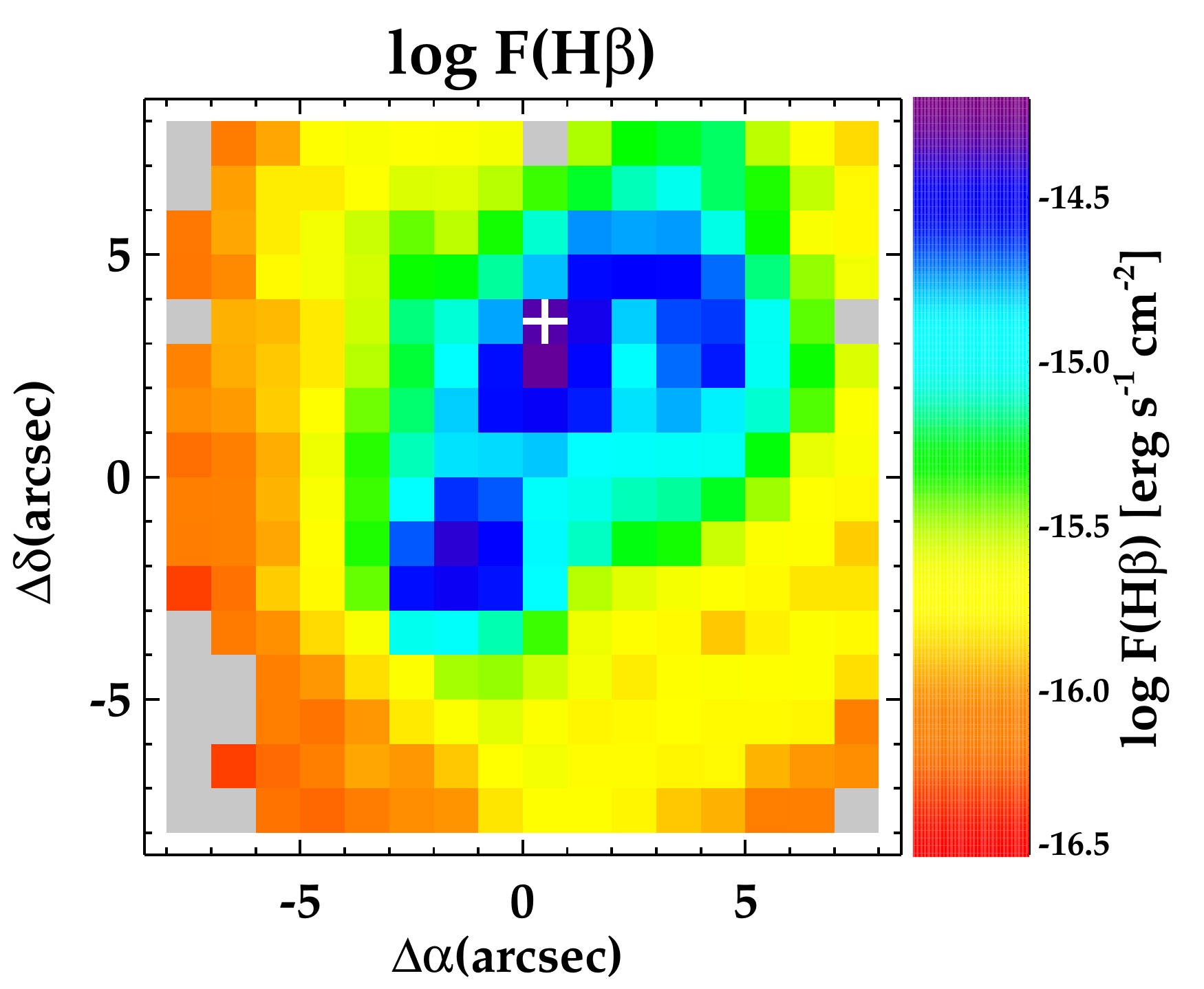}\\
\includegraphics[bb=1 1 505 405,width=0.45\textwidth,clip]{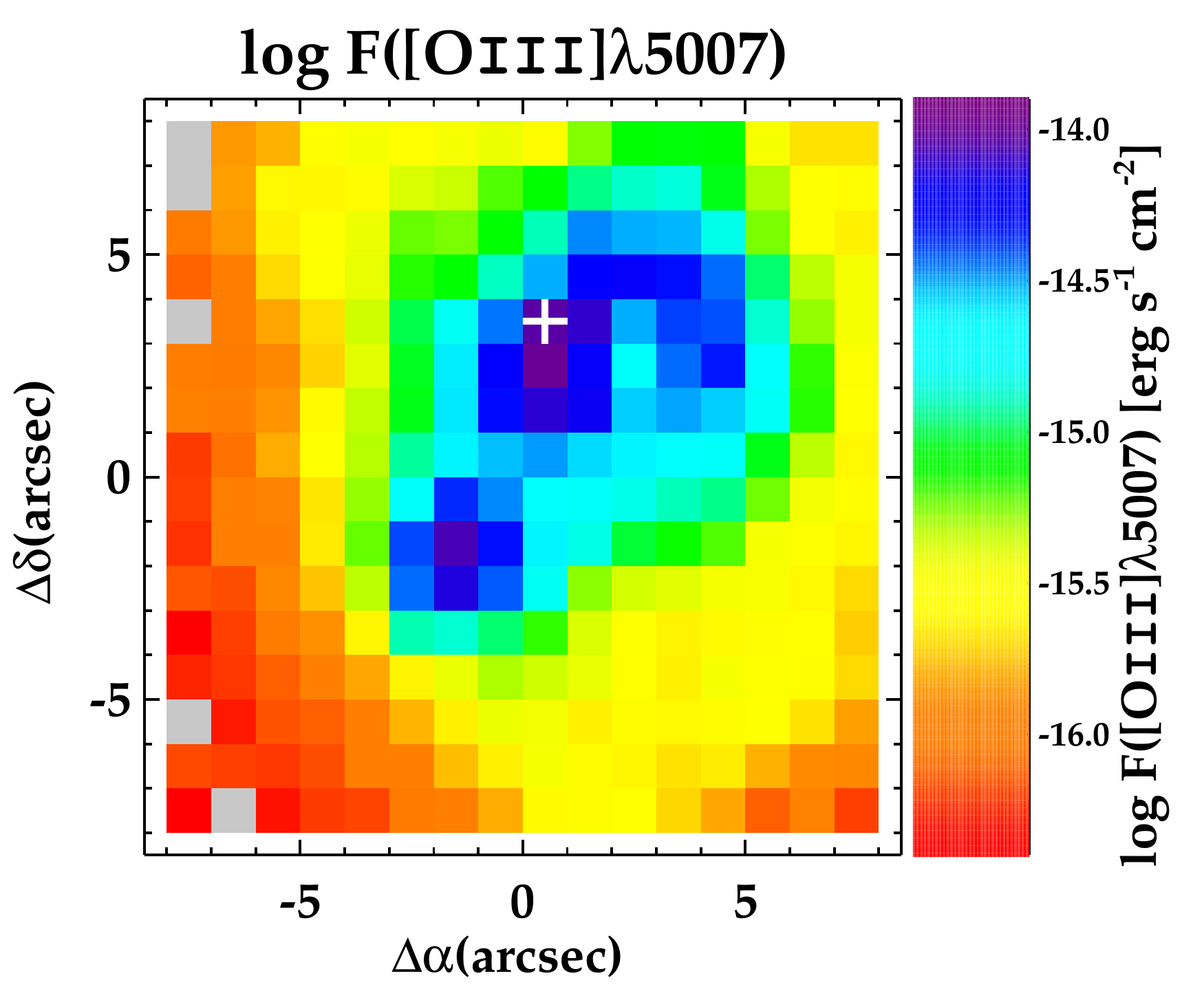}
\includegraphics[bb=1 1 505 405,width=0.45\textwidth,clip]{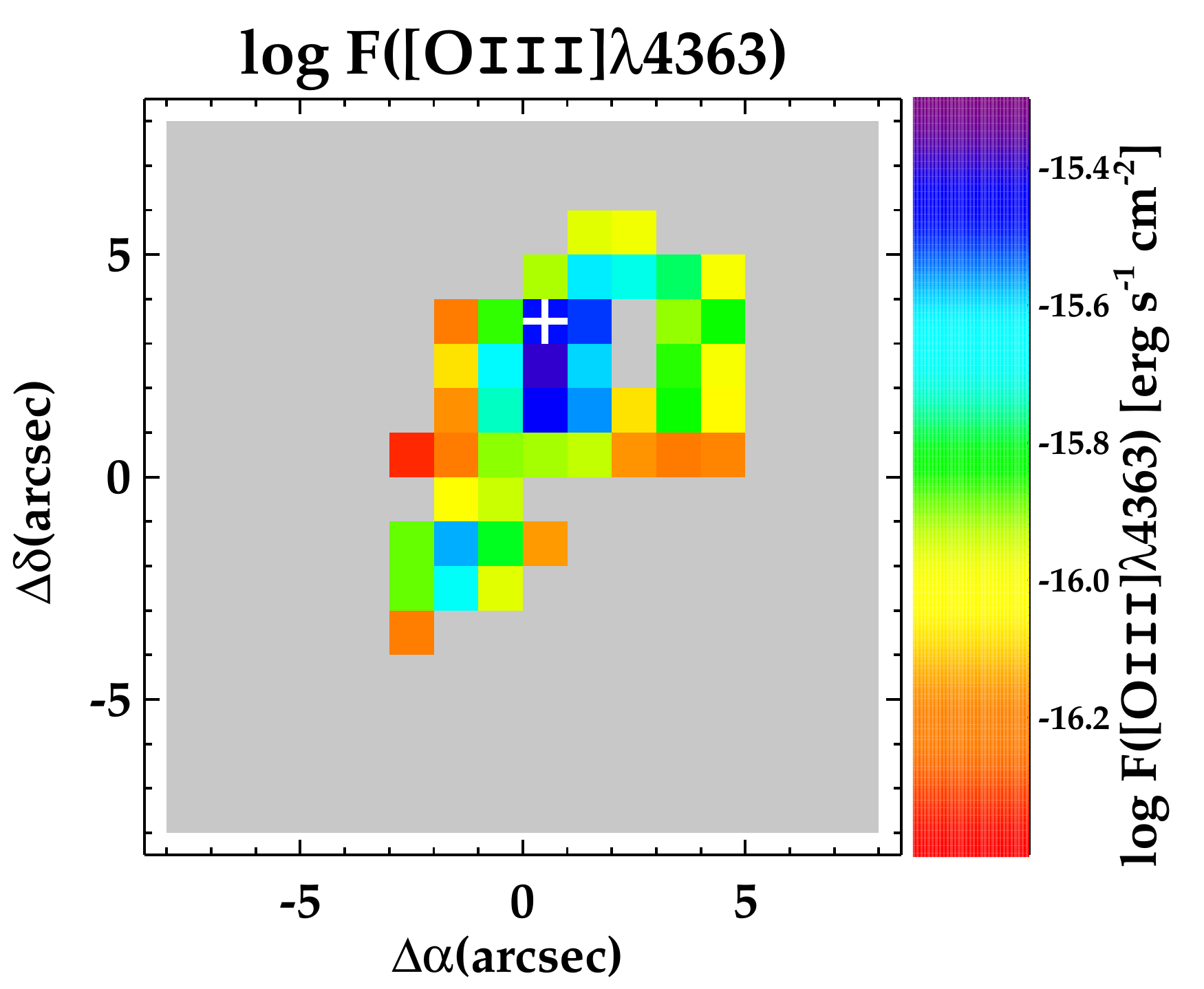} \\
\includegraphics[bb=1 1 505 405,width=0.45\textwidth,clip]{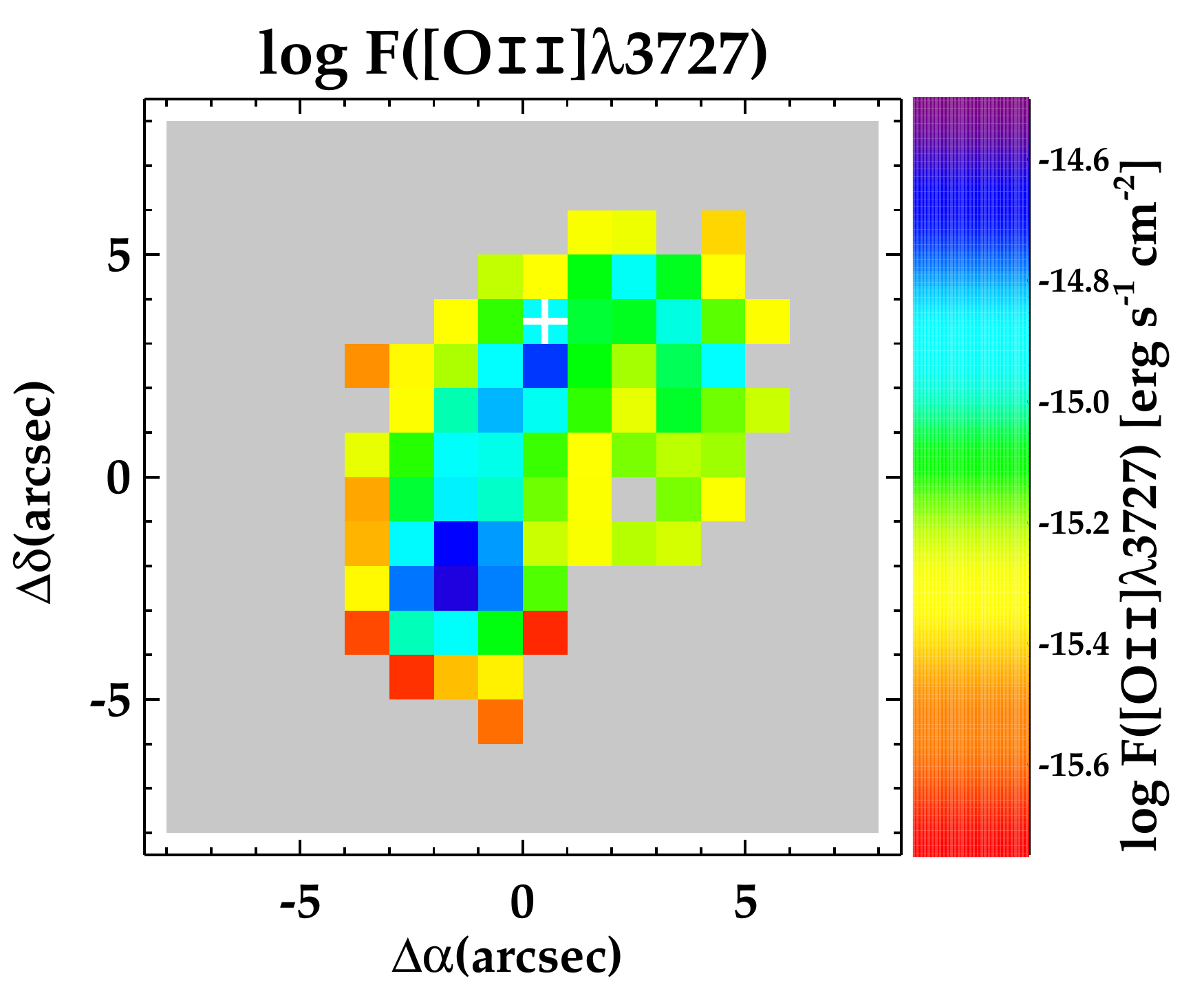}
\caption{Emission-line flux maps of IZw18: [O{\sc ii}]$\lambda$3727, H$\beta$,
  [O{\sc iii}]$\lambda$4363, [O{\sc iii}]$\lambda$5007, H$\alpha$. The
  spaxels with no measurements available are left grey. All maps are
  presented in logarithmic scale. As a guide to the reader, the peak
  of H$\alpha$ emission is marked with a plus (+) sign on all
  maps. The cross on the H$\alpha$ map marks the spaxel where we
  detect the WR feature (see Section~\ref{sris}).  The H$\alpha$ map also shows the boundaries
  of the areas that we use to create the integrated spectra of the NW
  and SE knots, and of the ``plume'' region (see text for
  details). North is up and east to the left.}
\label{line_maps} 
\end{figure*}

Using our own IDL scripts, we combine the line fluxes with the
position of the fibres on the sky to create the maps of emission lines
presented in this paper.  Fig.~\ref{line_maps} displays the [O{\sc
ii}]$\lambda$3727, H$\beta$, [O{\sc iii}]$\lambda$4363, [O{\sc
iii}]$\lambda$5007, and H$\alpha$ emission line maps. As a guide to
the reader, the spaxel\footnote{Individual elements of IFUs are often called ``spatial pixels'' (commonly shortened to ``spaxel''); the term is used to differentiate between a spatial element on the IFU and a pixel on the detector. } that corresponds to the H$\alpha$ emission peak is
indicated in all maps. The global structure of all the emission line
maps is similar, but not all display the same area over the IZw18
FOV. The intensity distribution of H$\beta$, [O{\sc
iii}]$\lambda$5007, and H$\alpha$ are more extended than that of
[O{\sc ii}]$\lambda$3727 and [O{\sc iii}]$\lambda$4363 because those
lines are among the brightest optical emission lines in the IZw18
spectra. In the H$\alpha$ map, we indicated the NW and SE knots, and
an arclike structure (called here ``plume'') which has one end rooted
in the vicinity of the NW knot.  From the maps of H$\beta$ and [O{\sc
iii}]$\lambda$5007, we can also distinguish these three regions.

The spatial distribution of the emission in [O{\sc iii}]$\lambda$4363,
  H$\beta$, [O{\sc iii}]$\lambda$5007 and H$\alpha$ are peaked on the
  NW component while the [O{\sc ii}]$\lambda$3727 emission reaches its
  maximum at the SE component. By inspecting the distribution of the
  [O{\sc iii}]$\lambda$5007 emission line, we can see that its peak
  intensity in the NW knot is slightly larger than that in the SE
  knot. In the case of the [O{\sc iii}]$\lambda$4363, H$\beta$ and
  H$\alpha$ maps, the ratio between the peak intensities from the NW
  knot and those from  the
  SE knot is $\sim$ 1.3-1.5, with the highest ratio found for the
  [O{\sc iii}]$\lambda$4363 map.

The reddening coefficient corresponding to each spaxel, c(H$\beta$),
was computed from the ratio of the measured-to-theoretical
H$\alpha$/H$\beta$ assuming the reddening law of \cite{cardelli89},
and case B recombination with electron temperature $T_{e}$ =
2$\times$10$^{4}$ K and electron density $n_{e}$ = 100 cm$^{-3}$ which
give an intrinsic value of H$\alpha$/H$\beta$= 2.75 \citep[][]{OF06}.

Typical values of the absorption H$\beta$ equivalent width, EW(H$\beta$)$_{abs}$,  found
for line-emitting SF galaxies are in the range $\sim$
0–-2 \AA~\citep[e.g.,][]{mc85,C09mrk409,EPM09}. Considering the high
values for EWs of Balmer emission lines measured from our data [e.g.,
EW(H$\beta$) $\sim $ 40-500 \AA], the effect of the underlying stellar
population in these lines appears not to be important. If we adopted an  
EW(H$\beta$)$_{abs}$ $\sim$ 2 \AA, the underlying absorption
correction would be typically small, less than 5$\%$ in H$\beta$.

\section{Spatially resolved ionization structure}
\label{sris}

\begin{figure*}
\center
\includegraphics[bb=1 1 505 405,width=0.45\textwidth,clip]{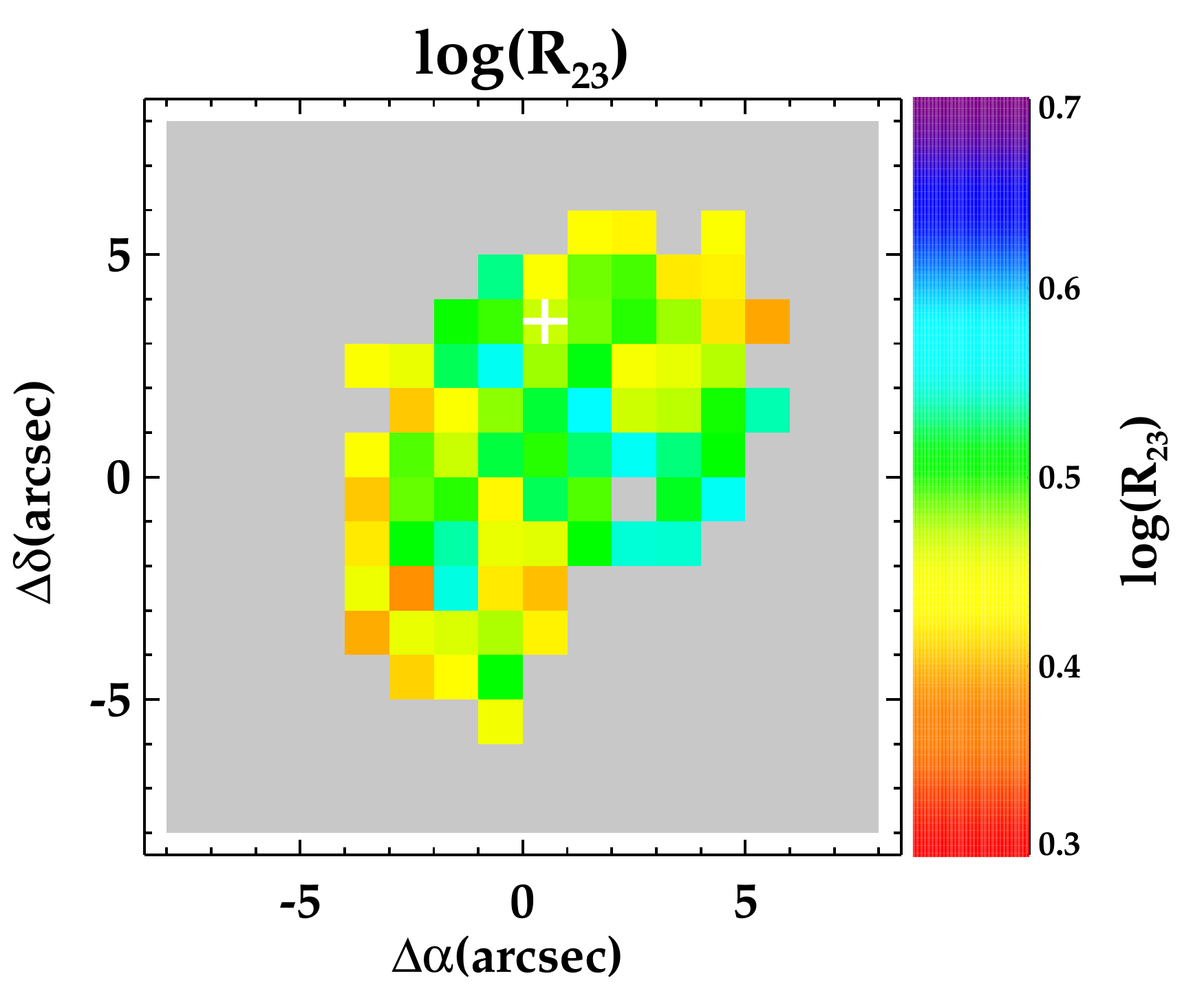}
\includegraphics[bb=1 1 505 405,width=0.45\textwidth,clip]{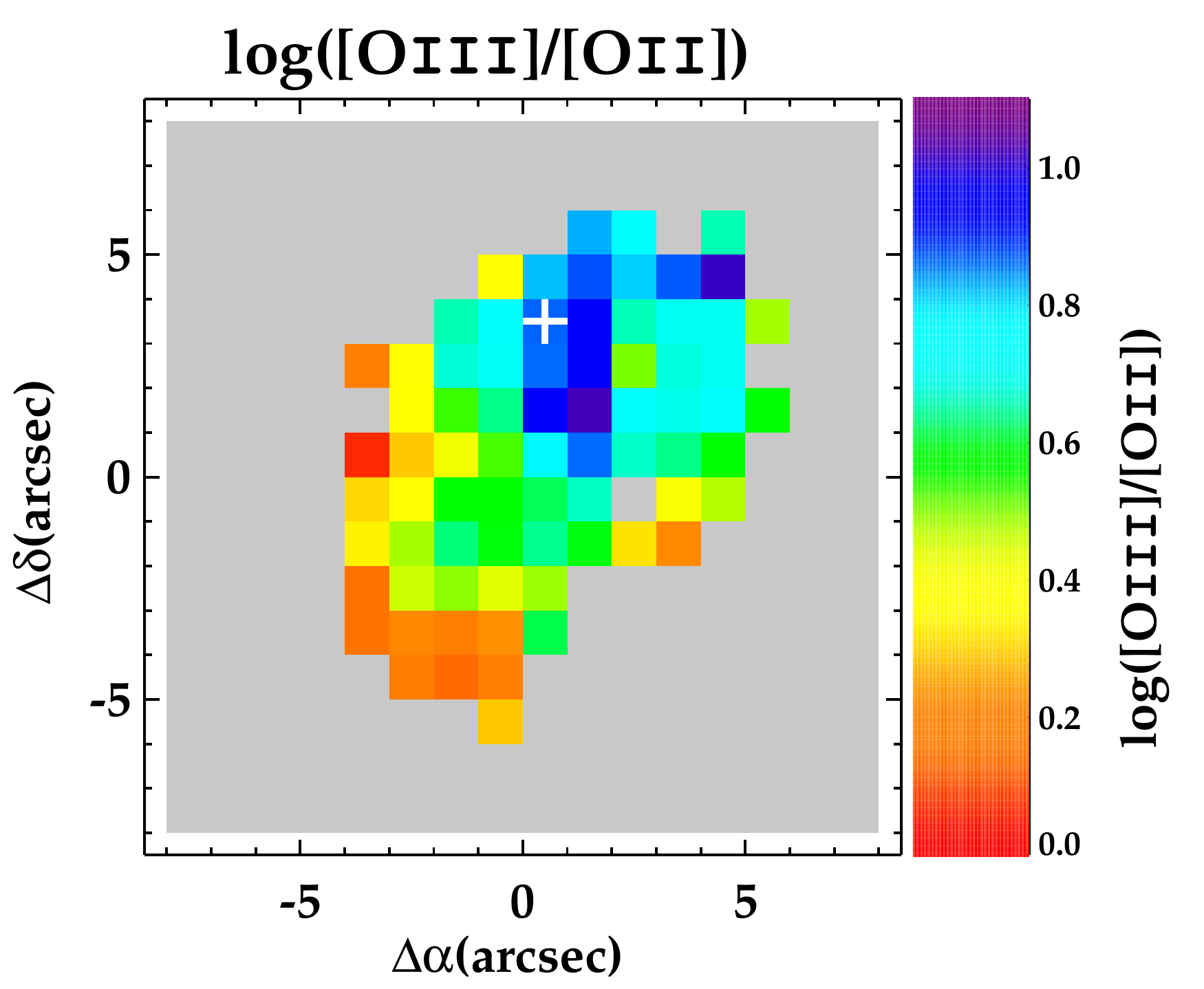}\\
\includegraphics[bb=1 1 505 405,width=0.45\textwidth,clip]{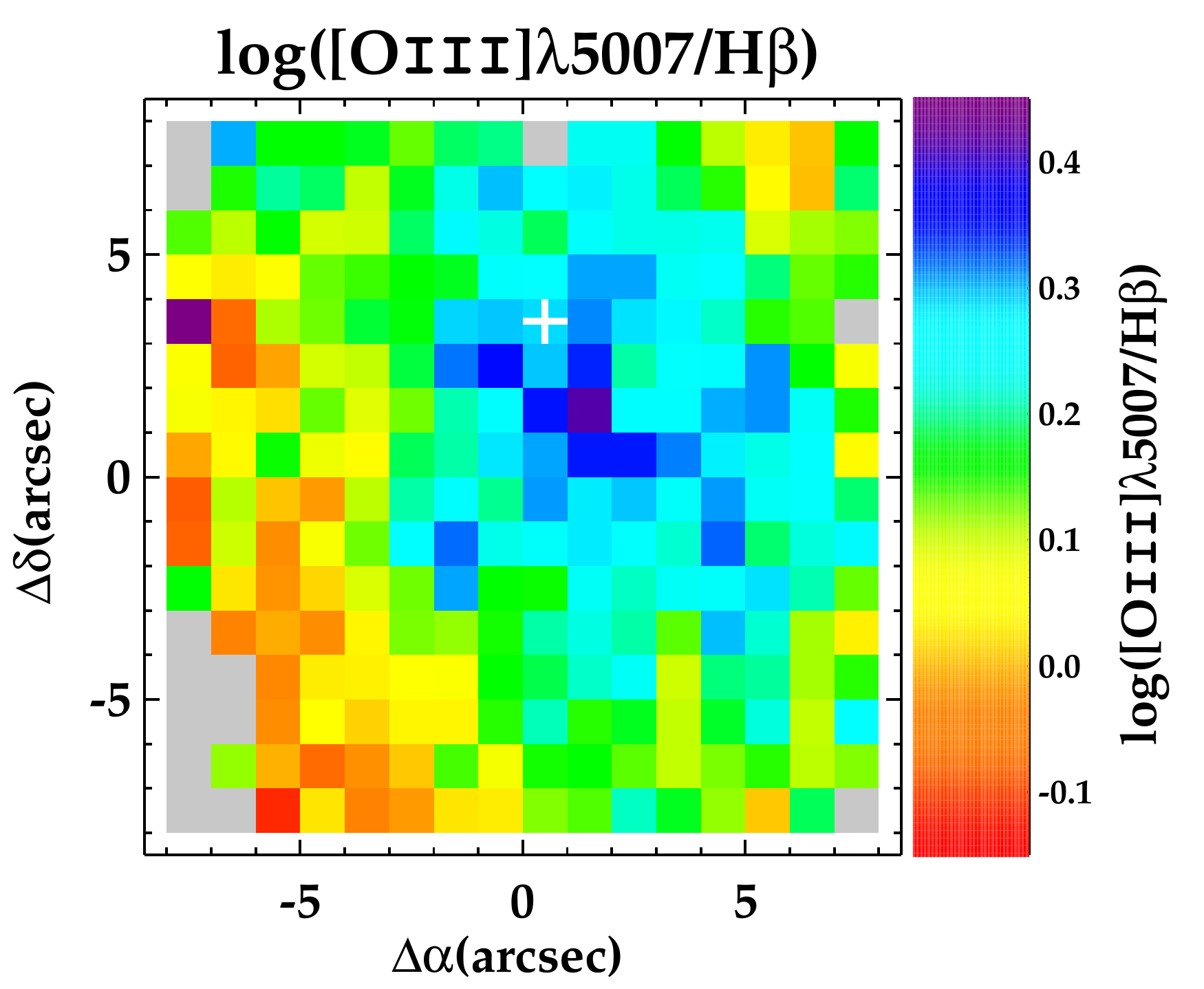}
\includegraphics[bb=1 1 505 405,width=0.45\textwidth,clip]{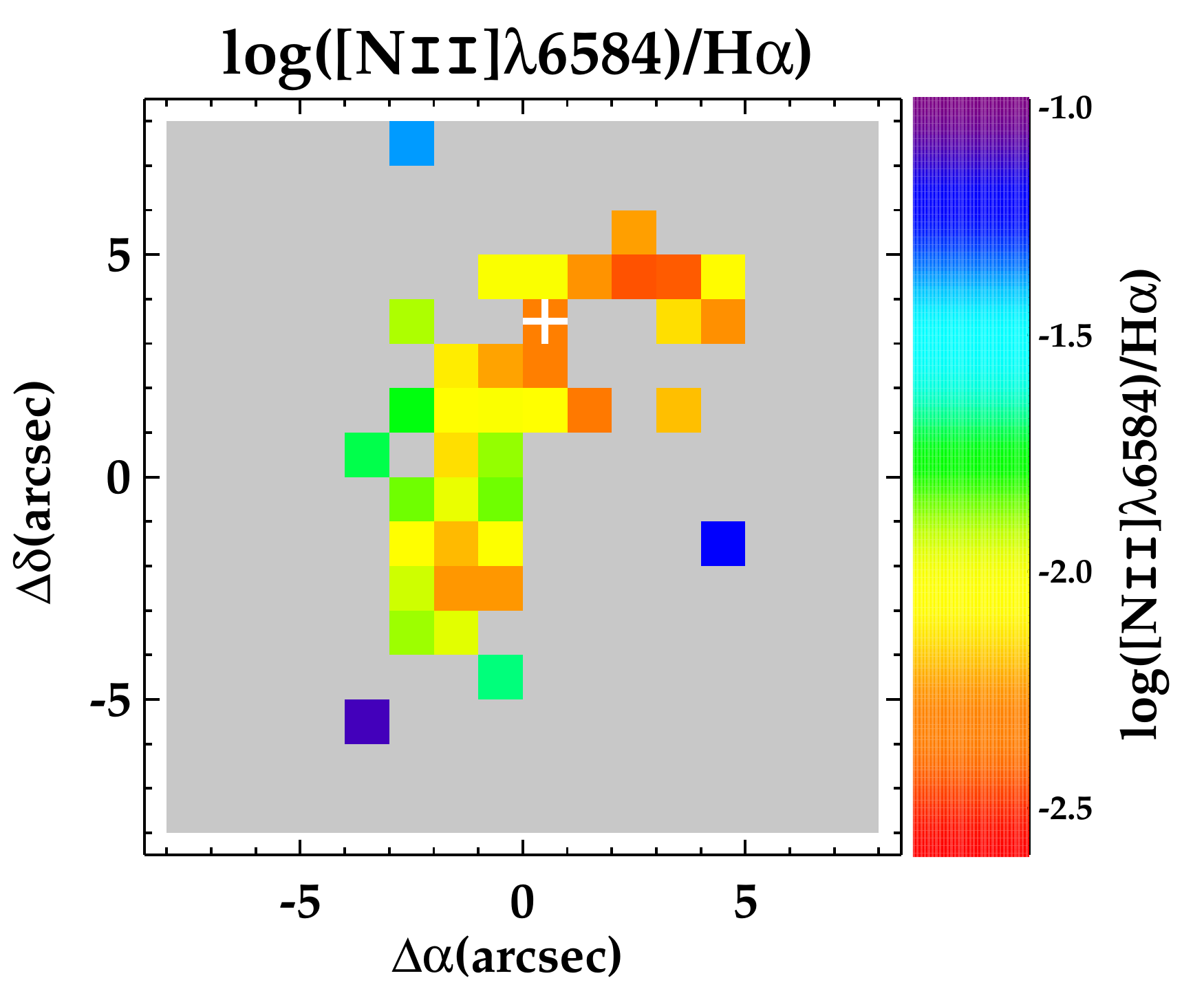}\\
\includegraphics[bb=1 1 505 405,width=0.45\textwidth,clip]{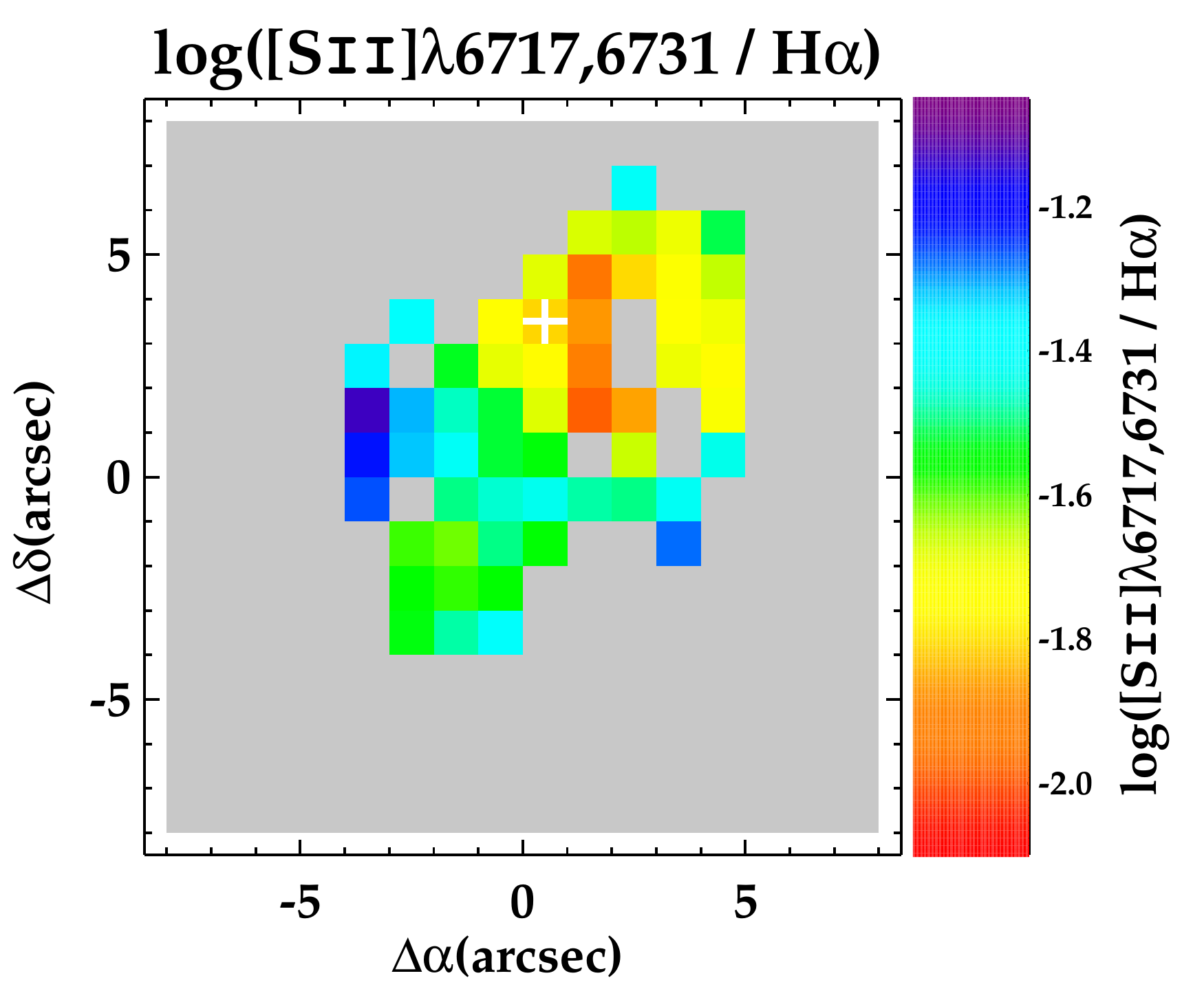}
\includegraphics[bb=1 1 505 405,width=0.45\textwidth,clip]{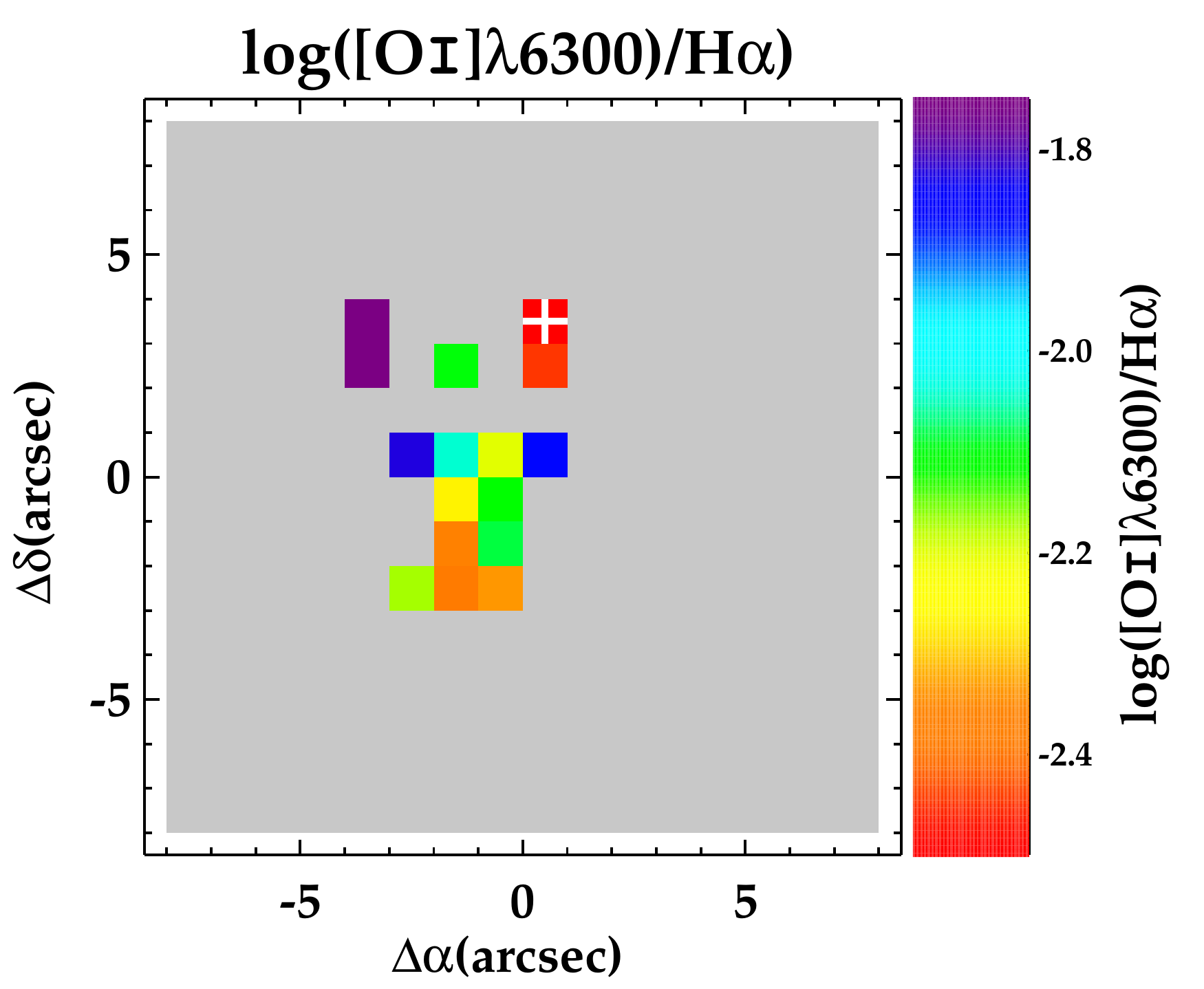}
\caption{Line ratio maps of IZw18: R$_{23}$, {\mbox [O{\sc
      iii}]/[O{\sc ii}]}, {\mbox [O{\sc
      iii}]$\lambda$5007/H$\beta$},  {\mbox [N{\sc
      ii}]$\lambda$6584/H$\alpha$,  {\mbox [S{\sc
        ii}]$\lambda$6717,6731/H$\alpha$},  {\mbox [O{\sc
        i}]$\lambda$6300/H$\alpha$}}. The spaxels with no measurements available are left
  grey. All maps are presented in logarithmic scale. As a guide to the
  reader, the peak of H$\alpha$ emission is marked with a plus (+) sign
  on all maps. North is up and east to the left.}
\label{line_ratio_maps} 
\end{figure*}

Maps of some of the relevant line ratios which are indicators of the ionization
structure for ionized gaseous nebulae are displayed in
Fig.~\ref{line_ratio_maps}. These line ratios are corrected for
reddening using the corresponding c(H$\beta$) for each spaxel.

\begin{figure}
\includegraphics[width=9.0cm,clip]{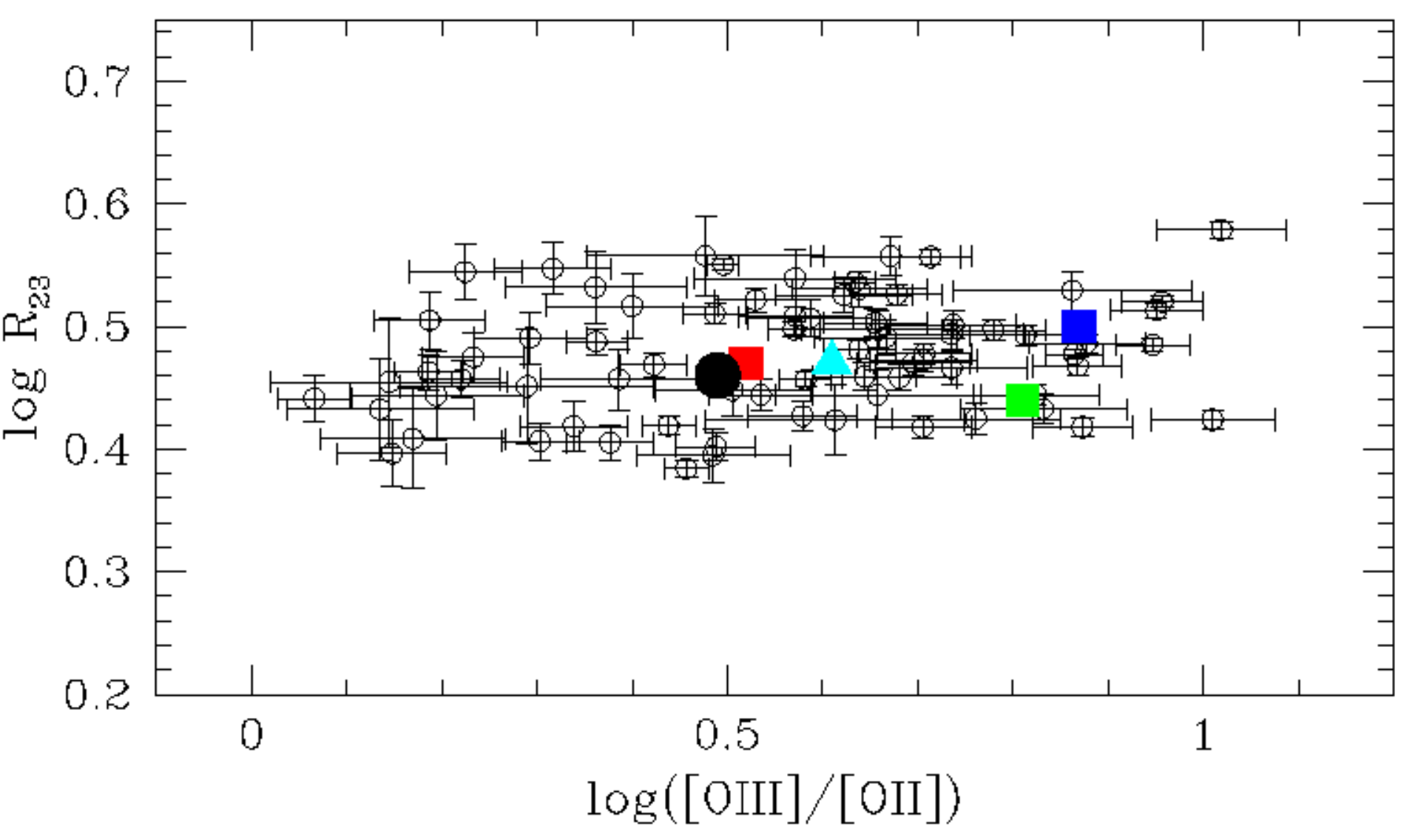}
\caption{The relation between log R$_{23}$ and log([O{\sc iii}]/[O{\sc ii}]). Open circles correspond to individual spaxels from the datacube. Overplotted as blue, red, and green squares are the line ratios measured from the 1D spectra of the NW knot, SE knot and ``plume'', respectively; the cyan triangle shows the line ratio values from the total integrated spectrum of IZw18; the black circle corresponds to the line ratios from the spectrum of the ``halo'' of IZw18 (see Section~\ref{int} and Table~\ref{table_regions} for details on the 1D-spectra extracted for selected galaxy regions).}
\label{z_u} 
\end{figure}

The spatial distribution of the abundance indicator R$_{23}$=([O{\sc
ii}]$\lambda$3727+[O{\sc iii}]$\lambda\lambda$4959,5007)/H$\beta$, first introduced by \citet[][]{pagel79}
and afterwards calibrated by several
authors \citep[e.g.,][]{mc85,mg91}, is found to be relatively flat
without any significant peak (see the map of R$_{23}$ in
Fig.~\ref{line_ratio_maps}).  However, the commonly-used ionization
parameter diagnostic [O{\sc iii}]/[O{\sc ii}]\footnote{[O{\sc iii}]/[O{\sc ii}]=(4959+5007)/3727} does not show {\bf a}
homogeneous spatial distribution with the highest values of [O{\sc
iii}]/[O{\sc ii}] found within the NW knot (see the corresponding map
in Fig.~\ref{line_ratio_maps}). Fig.~\ref{z_u} shows the relation
between the measured values of R$_{23}$ and [O{\sc iii}]/[O{\sc ii}]
from which it is clear that there is no dependence between R$_{23}$
and the ionization parameter (as traced by [O{\sc iii}]/[O{\sc ii}])
across IZw18 \citep[see also][]{jvm98}. While R$_{23}$ remains
practically constant (0.40 $\leq$ log R$_{23}$ $\leq$ 0.55), [O{\sc
iii}]/[O{\sc ii}] presents changes larger than a factor of 8 across
the FOV. The observations of the giant HII region
NGC~604 by \citet[][]{ngc604} give another
example where a large range of excitation is seen whereas the
metallicity indicator R$_{23}$ remains substantially
constant \citep[see also][]{P11}. In Section~\ref{stat} we will discuss the
observed variations of R$_{23}$ and [O{\sc iii}]/[O{\sc ii}] from a
statistical point of view.

\begin{figure} 
  \centering
\includegraphics[width=9.0cm,clip]{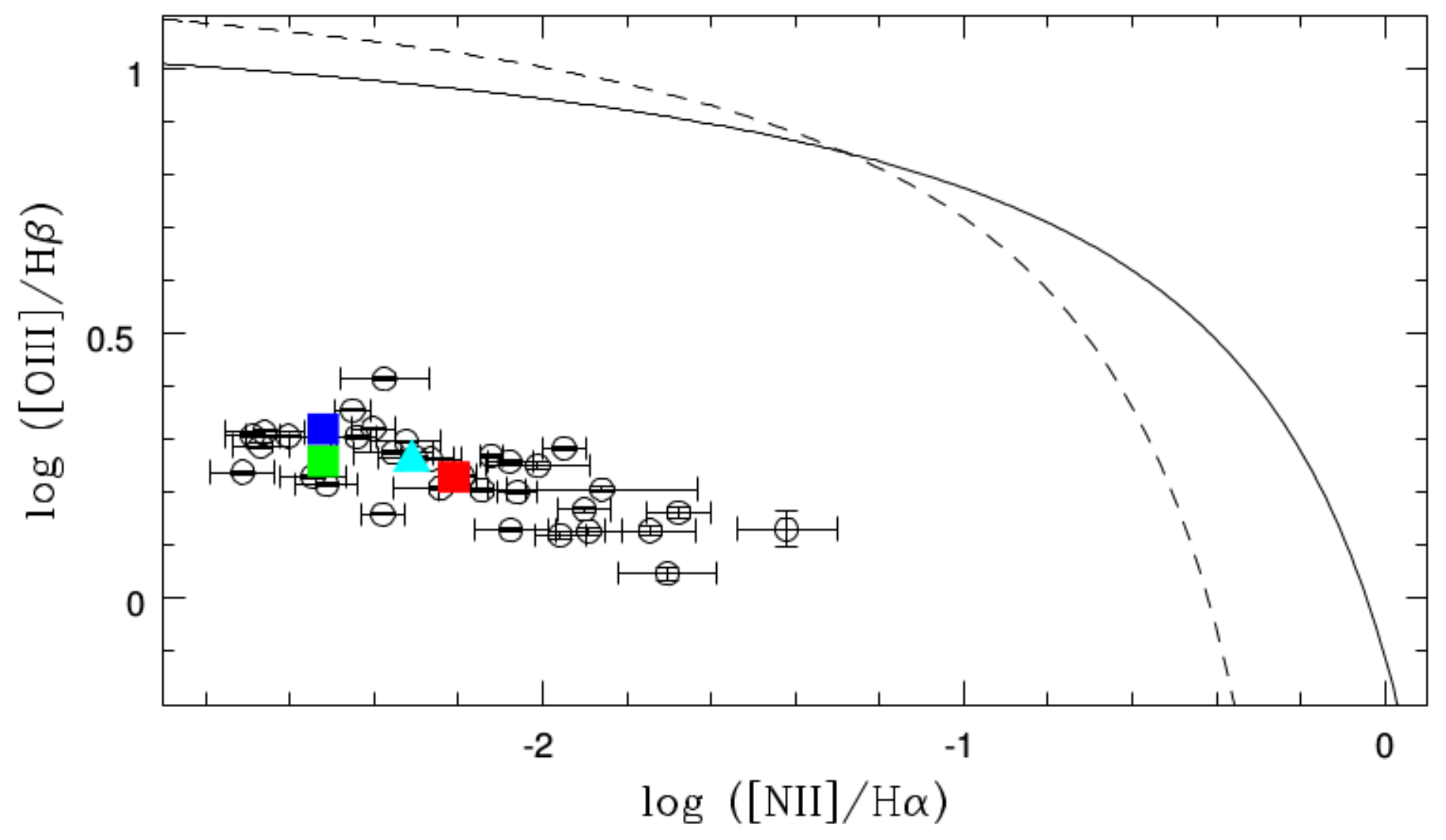} 
\includegraphics[width=9.0cm,clip]{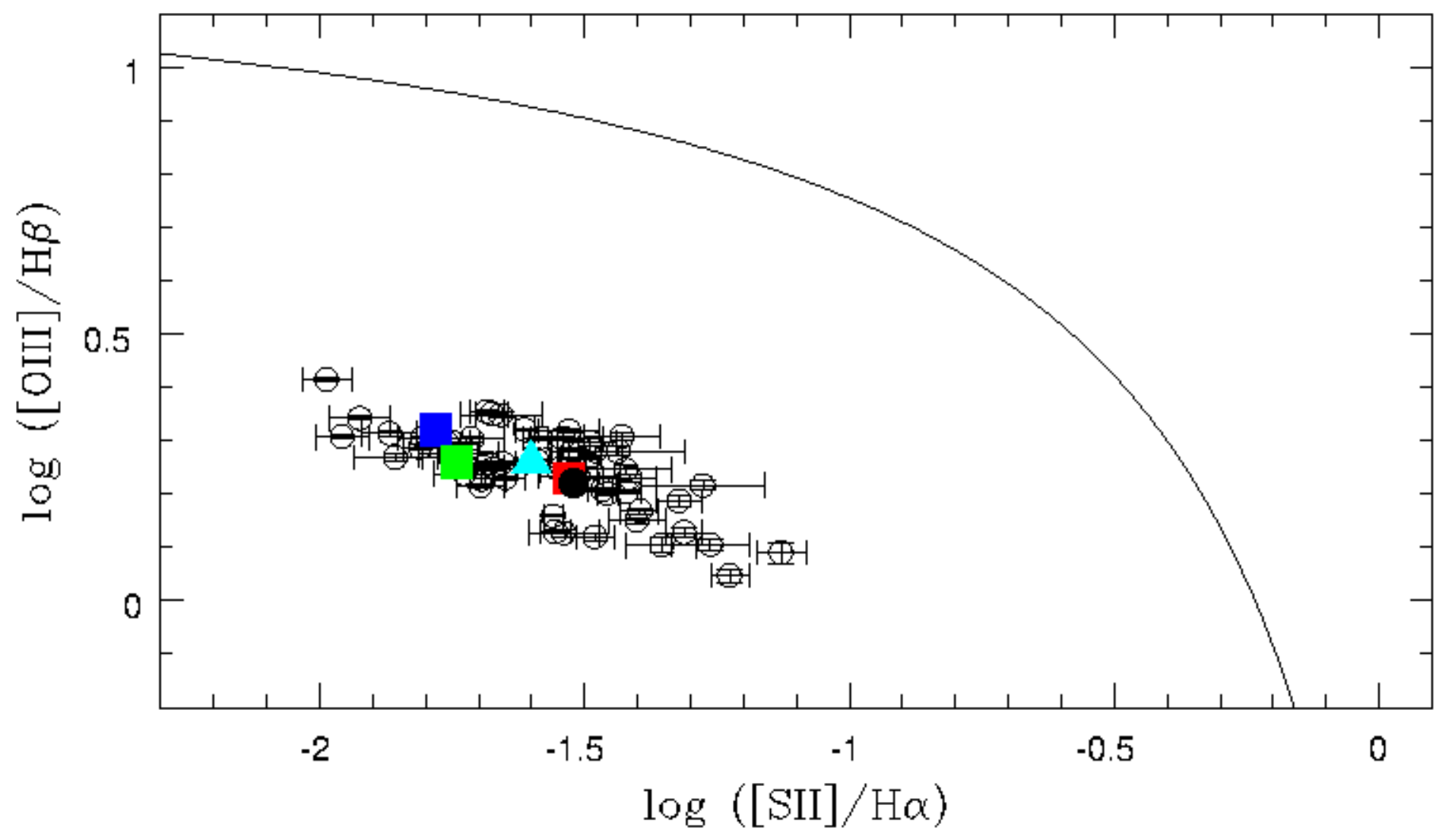} 
\includegraphics[width=9.0cm,clip]{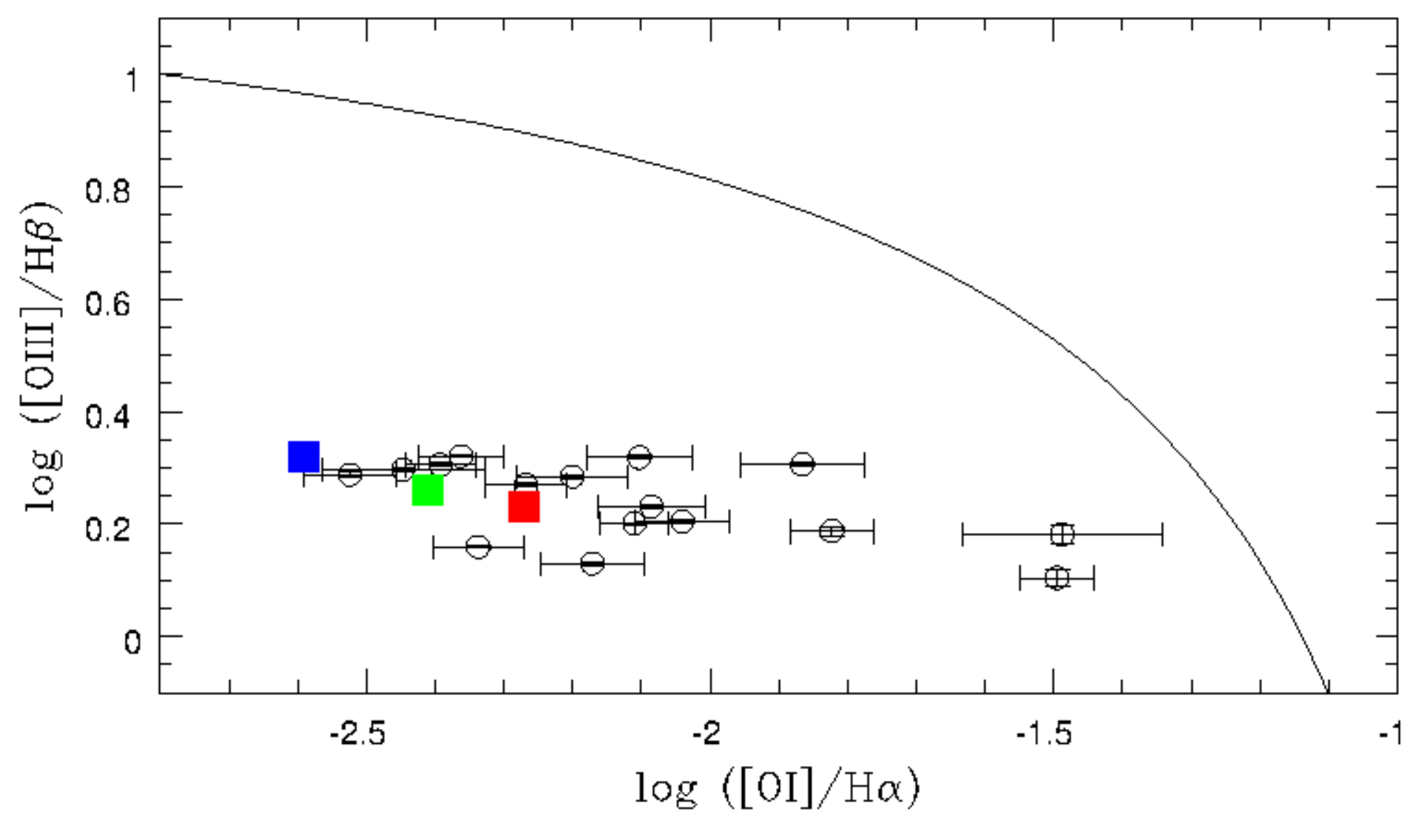}
\caption{BPT diagnostic diagrams for IZw18. From top to bottom: log([O{\sc iii}]$\lambda$5007/H$\beta$) vs. log ([N{\sc ii}]$\lambda$6584/H$\alpha$), log ([O{\sc iii}]$\lambda$5007/H$\beta$) vs. log ([S{\sc ii}]$\lambda$6731,6717/H$\alpha$) and log ([O{\sc iii}]$\lambda$5007/H$\beta$) vs. log ([O{\sc i}]$\lambda$6300/H$\alpha$). The symbols are as described in Fig. 4. Overplotted as a black solid curve (in all three panels) is the theoretical maximum starburst model from \citet{kew01}, devised to isolate objects whose emission line ratios can be accounted for by the photoionization by massive stars (below and to the left of the curve) from those where some other source of ionization is required. The black-dashed curve in the [N{\sc ii}]$\lambda$6584/H$\alpha$ diagram represents the demarcation between SF galaxies (below and to the left of the curve) and AGNs defined by \citet{K03}.}
\label{bpt} 
\end{figure}

In the maps shown in Fig.~\ref{line_ratio_maps}, high values of [O{\sc
iii}]/[O{\sc ii}] and [O{\sc
iii}]$\lambda$5007/H$\beta$, and low values of [S{\sc
ii}]$\lambda\lambda$6717,6731/H$\alpha$ and [N{\sc
ii}]$\lambda$6584/H$\alpha$ correspond to the areas of ionized gas
with relatively high excitation. By inspecting these maps, there is a
clear tendency for the gas excitation to be higher at the location of
the NW knot and thereabouts, in comparison to the SE component.
Here we should also note that \citet[][]{K15} found the NW component to be very close to the 
nebular
HeII$\lambda$4686-emitting region \citep[see Fig. 2 from][]{K15}. All this indicates the presence of a harder
ionizing field in the NW knot of IZw18. We will discuss more
about it in Section~\ref{2dgas}. Another feature of the ionization structure
in IZw18 is the existence of a relatively high-excitation diffuse gas
outside of the main SF knots indicated by an extended low
surface brightness emission in the maps of [O{\sc
iii}]$\lambda$5007/H$\beta$.

The standard diagnostic diagrams \citep[][hereafter BPT]{bal81} are a
powerful tool, widely used to identify the dominant mechanism of gas
excitation, i.e. either photoionization by massive stars within HII
regions or other ionizing sources, including photoionization by
AGNs, post-AGB stars and shocks \citep[e.g.,][]{K12,S15,G15}. The BPT diagrams, on a spaxel-by-spaxel basis,
for IZw18 are shown in Fig.~\ref{bpt}: [O{\sc
iii}]$\lambda$5007/H$\beta$ vs. [N{\sc ii}]$\lambda$6584/H$\alpha$,
[S{\sc ii}]$\lambda\lambda$6717,6731/H$\alpha$, and [O{\sc
i}]$\lambda$6300/H$\alpha$. The line ratios obtained from the
one-dimensional (1D) spectra of selected regions across our FOV (see
Section~\ref{int} and Table~\ref{table_regions}) are overplotted on
the BPT diagrams. For all positions in IZw18 our emission-line ratios
fall in the general locus of SF objects according to the
spectral classification scheme proposed by \cite{kew01}
and \cite{K03}, as indicated in Fig.~\ref{bpt}. This suggests that
photoionization from hot massive stars is the dominant excitation
mechanism within IZw18.

\begin{figure} 
\centering 
\includegraphics[width=7cm]{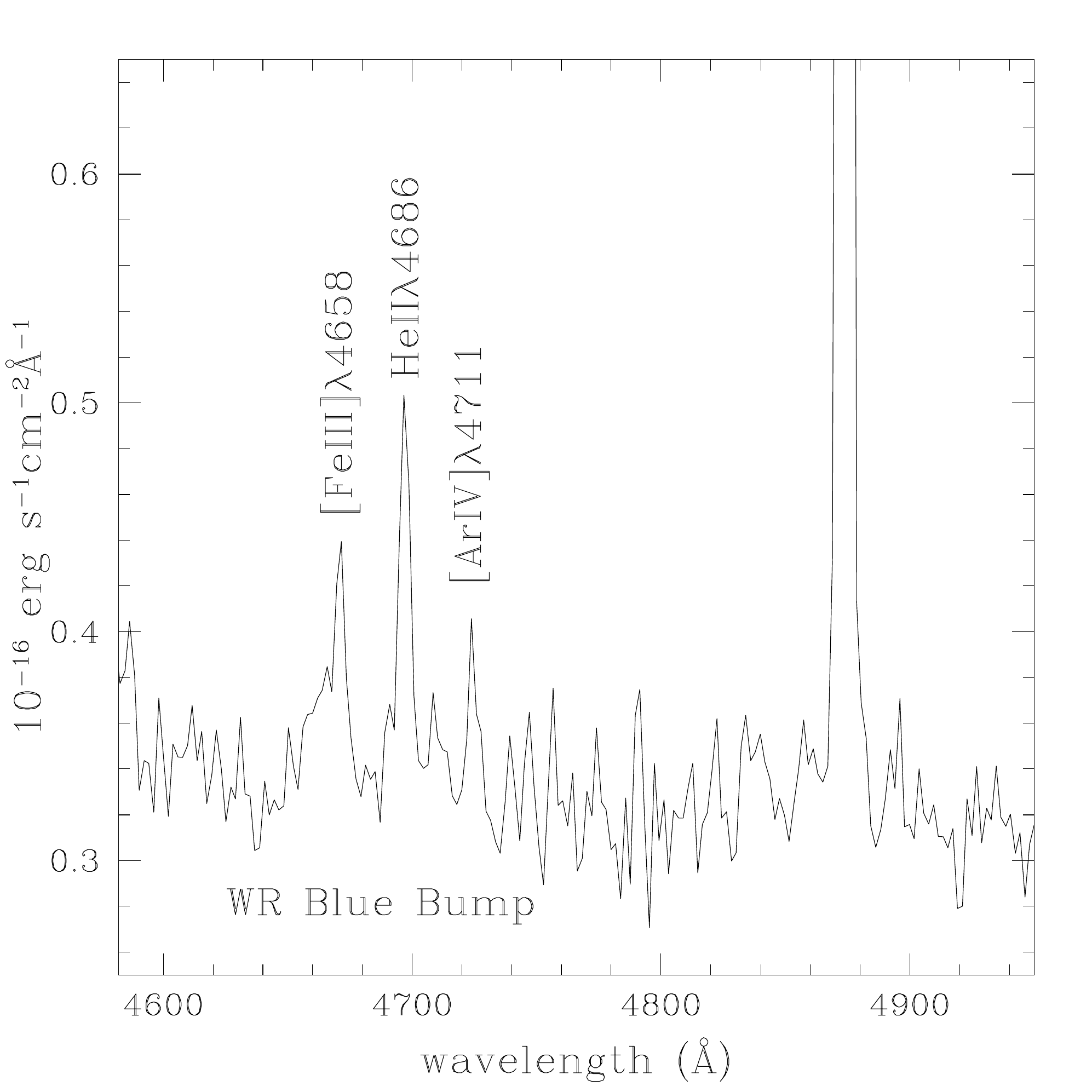}
\caption{Spectrum showing signs of a faint, broad underlying emission feature around $\sim$ 4650-4730 \AA, i.e., the WR blue bump. The location of the WR feature within our FOV
is indicated in the H$\alpha$ map from Fig.~\ref{line_maps}. The narrow [Fe{\sc iii}]$\lambda$4658, He{\sc ii}$\lambda$4686, and [Ar{\sc iv}]$\lambda$4711 emission lines are marked.}
\label{wr_bump} 
\end{figure}

Regarding the massive stellar content of IZw18, previous optical
long-slit spectroscopy detected Wolf-Rayet (WR) features (most
commonly a broad feature centred at $\sim$ 4680 \AA~ or ``blue bump'') in
the NW region \citep[e.g.,][]{L97,I97}. However the uncertainty on the
exact location of the slits has prevented these works to give the WR
star's position more precisely. Our data set show indications  of a weak WR
blue bump around $\sim$ 4650-4730 \AA~in one spaxel (see Fig.~\ref{wr_bump}), which lies in
the NW knot; more specifically it is
found to be $\sim$ 1.5$\arcsec$ ($\sim$ 130 pc at the distance of
IZw18) southeast of the H$\alpha$ peak (see the
H$\alpha$ flux map in Fig.~\ref{line_maps}). We cannot discard the
presence of even fainter WR features at other locations across our FOV
which might not be detected due to the signal-to-noise ratio of our
spectra. The observation of WRs in very metal-deficient objects, like
IZw18, keeps challenging current stellar evolutionary models for
single massive stars, which do not predict
any WRs in metal-poor environments \citep[e.g.,][]{L14}. Further investigation on formation channels for such metal-poor WRs is needed but is beyond the scope of this work \citep[see e.g.,][and references therein]{PC07}.

\section{A 2D view of the physical-chemical properties for the ionized gas}\label{2dgas}

\begin{figure*}
\center
\includegraphics[bb=1 1 505 405,width=0.45\textwidth,clip]{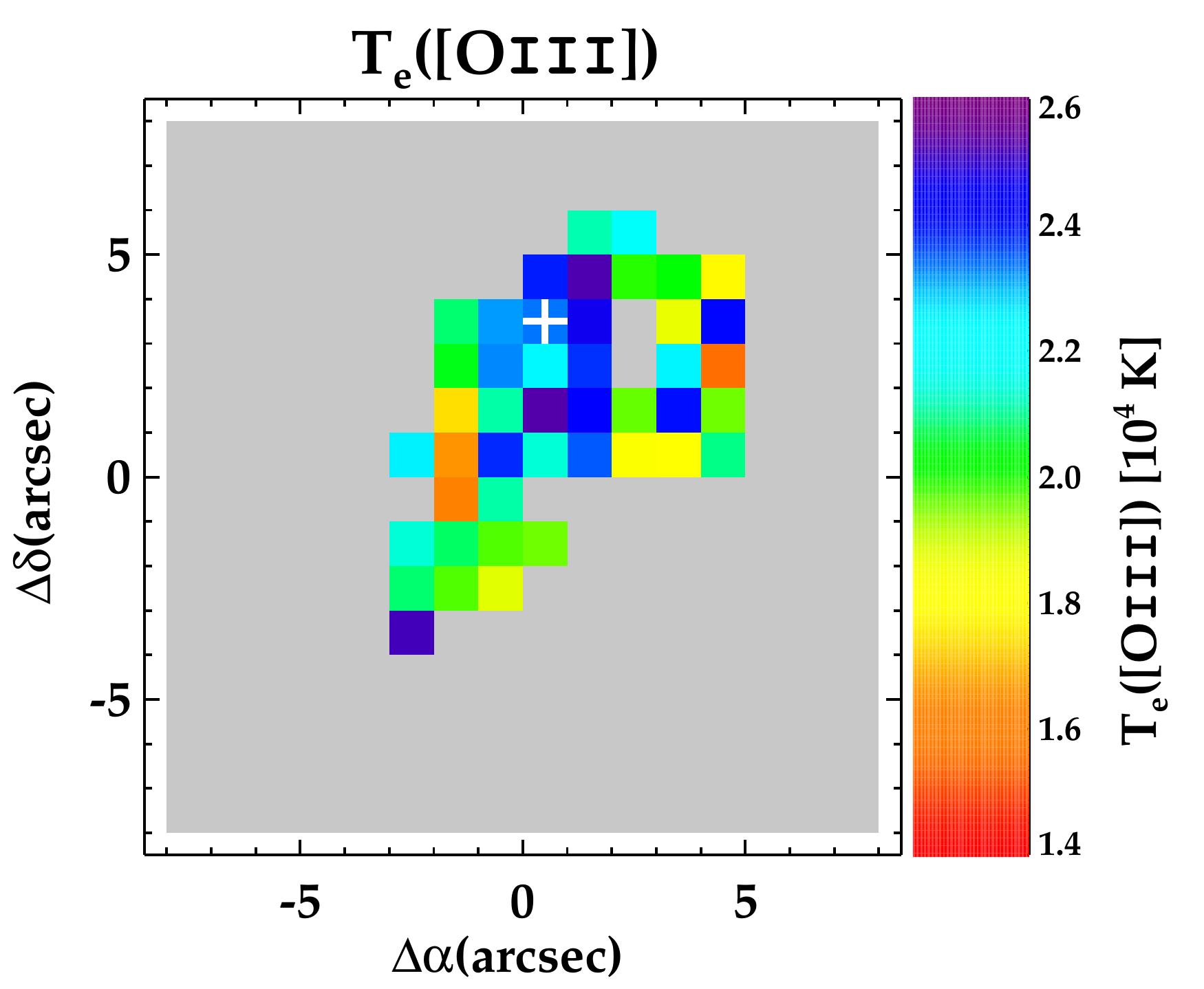}
\includegraphics[width=9.5cm,clip]{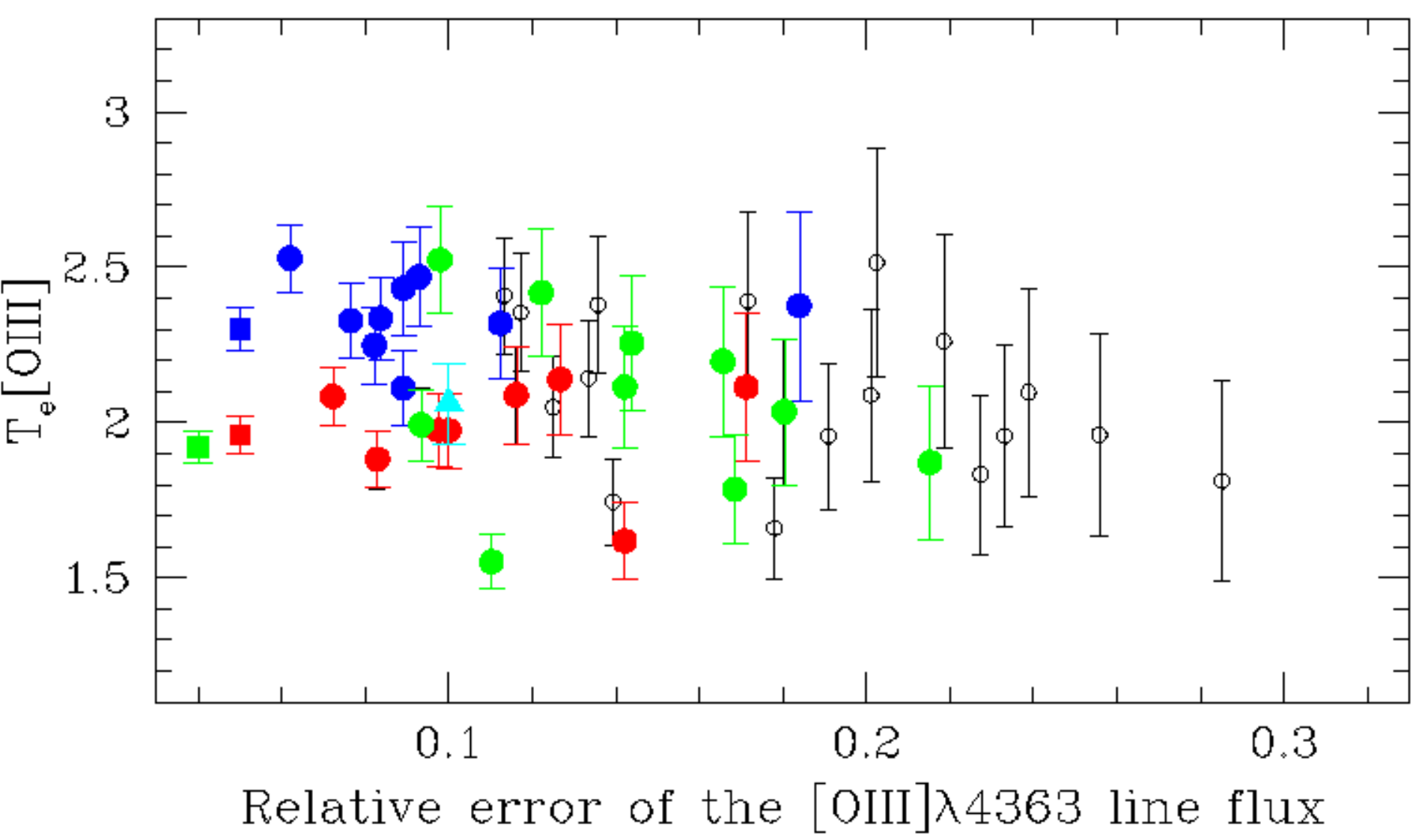}
\caption{{\it Left panel}: map of T$_{e}$[O{\sc iii}] derived directly
  from the measurement of the  [O{\sc iii}]$\lambda$4363 line flux. The spaxels with no measurements available are left
  grey. The peak of H$\alpha$ emission is marked with a plus (+) sign
  for orientation. North is up and east to the left. {\it Right
    panel}: T$_{e}$[O{\sc iii}] derived directly from
    the measurement of the [O{\sc
      iii}]$\lambda$4363 line flux versus the relative error of the
    measurement.  Open circles represent individual spaxels; blue, red
    and green circles indicate the individual spaxels used to create
    the 1D spectra of the NW knot, SE knot, and plume,
    respectively. Squares indicate the values measured from the 1D
    integrated spectra with the same colour-code as used for the
    individual spaxels. The cyan triangle shows the value measured
    from the IZw18 integrated spectrum (see Section~\ref{int} and Table~\ref{table_regions} for details on the 1D-spectra extracted for selected galaxy regions).}
\label{te} 
\end{figure*}

\begin{figure*}
\center
\includegraphics[bb=1 1 505 405,width=0.45\textwidth,clip]{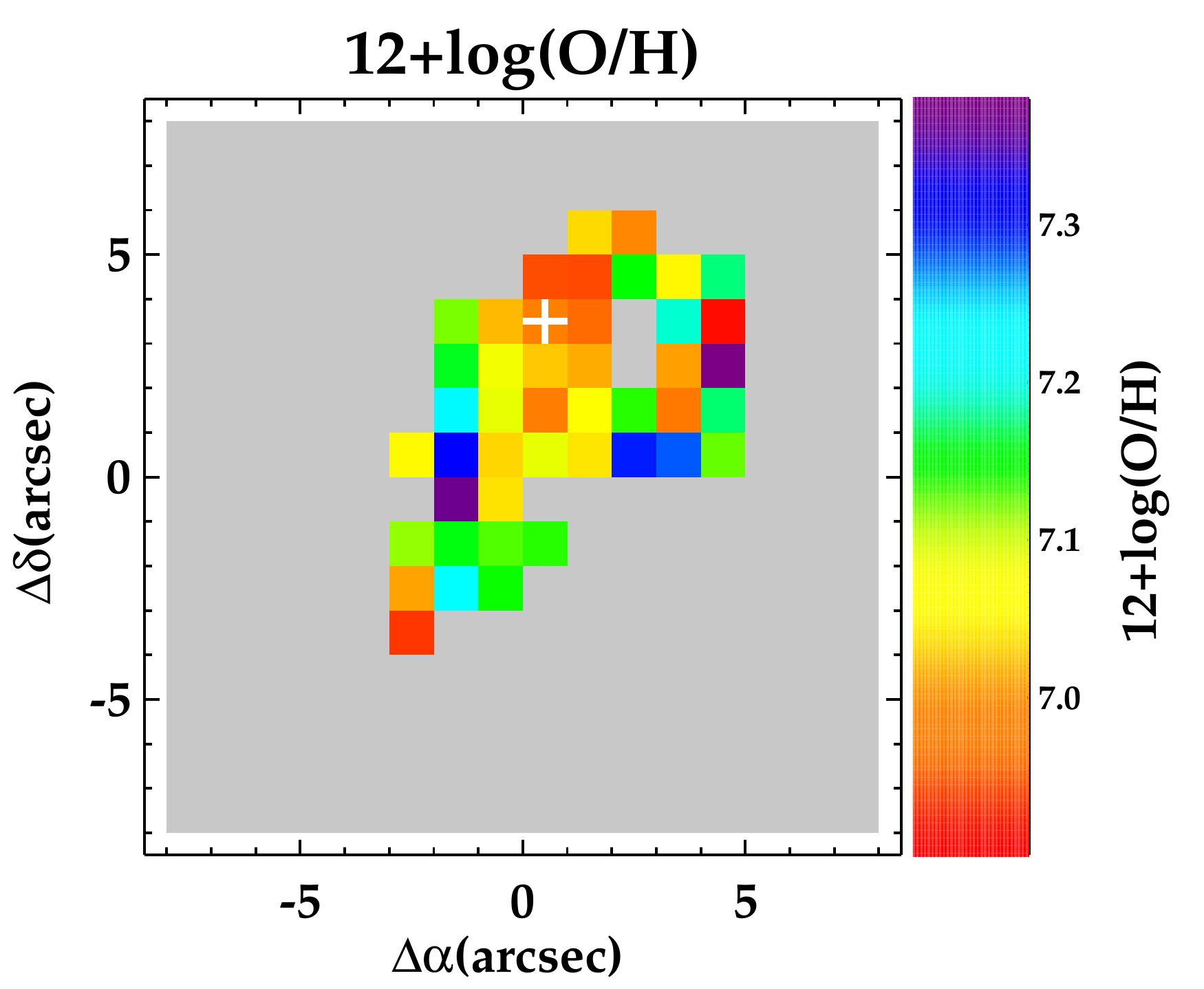}
\includegraphics[width=0.35\textwidth,clip]{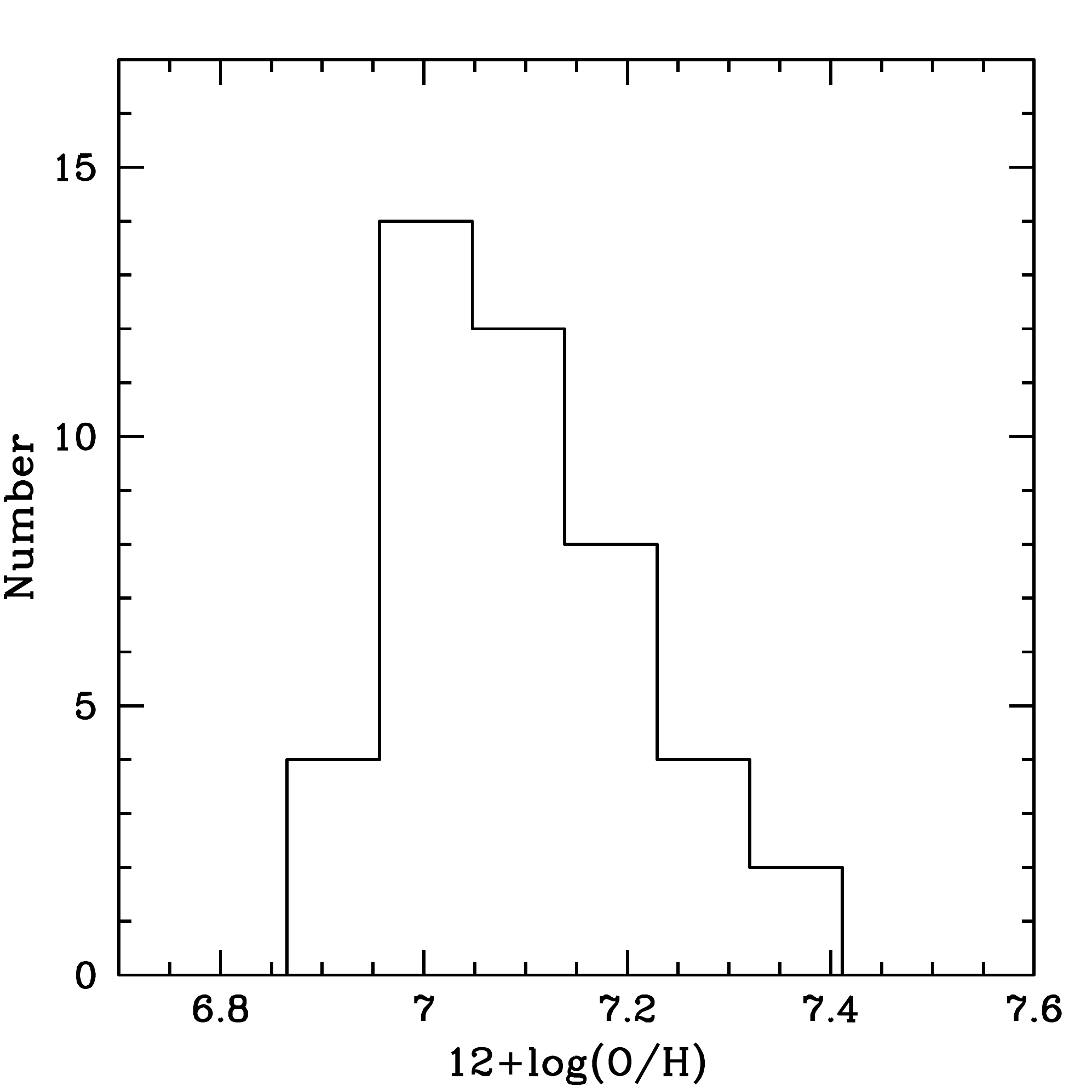}
\caption{{\it Left panel}: map of oxygen abundance (12+log O/H)
  derived from $T_{e}$[O{\sc iii}]. The corresponding histogram is in
  the {\it right panel}.}
\label{oh} 
\end{figure*}

\begin{figure} 
\centering 
\includegraphics[width=7.0cm,clip]{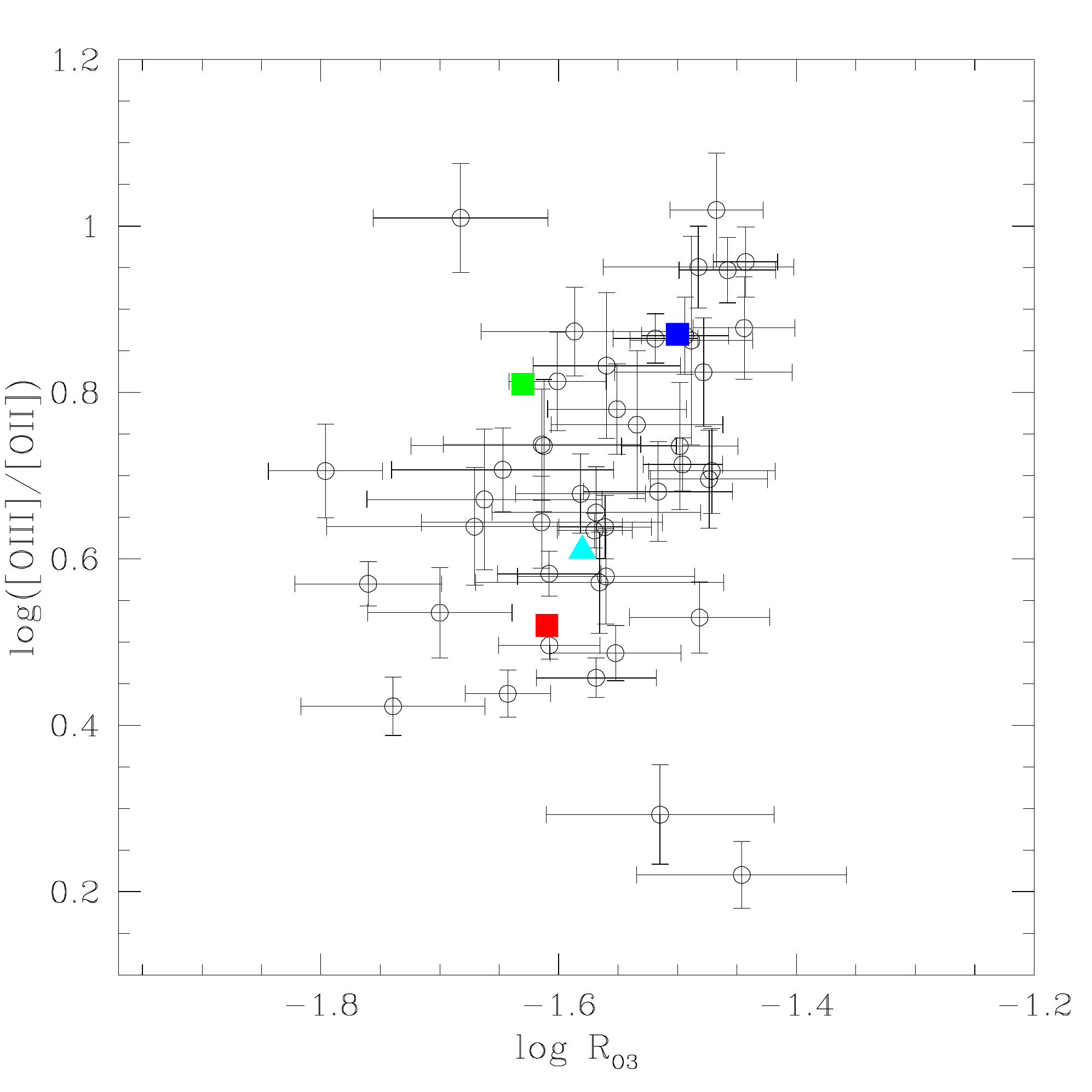}
\caption{log ([O{\sc iii}]]/[O{\sc ii}]) versus log R$_{O3}$ \mbox{[= log ([O{\sc
    iii}]$\lambda$4363/[O{\sc iii}]$\lambda$4959,5007)]}, i.e., the relation between the line-ratio indicators of the ionization parameter and electron temperature. The
  symbols are as described in Fig. 4.}
\label{o3o2_ro3} 
\end{figure}

In order to derive the physical properties and ionic abundances of the
ionized gas for IZw18, we have used the expressions given
by \citet{EPM14} which are obtained from the code
PYNEB \citep{L15}. We have calculated the final errors in the derived
quantities by error propagation and taking into account errors in flux
measurements. 

For more than 80$\%$ of the spectra where we measure the [S{\sc
ii}]$\lambda$6717/$\lambda$6731 line ratio, the derived $n_{\rm e}$
values are $\lesssim$ 300 cm$^{-3}$, which place most of spaxel
spectra in the low-density regime.

For the [O{\sc iii}]$\lambda$4363-emitting spaxels, we have derived the
$T_{\rm e}$ values of [O{\sc iii}] using the [O{\sc
  iii}]$\lambda$4363/[O{\sc iii}]$\lambda$4959,5007 line ratio,
corrected for extinction.  We were able to measure the faint auroral line
[O{\sc iii}]$\lambda$4363 above the 3$\sigma$ detection limit for 44
spaxels. These spaxels cover a projected area of nearly 42
arcsec$^{2}$ equivalent to 0.32 kpc$^{2}$, including the NW knot and
a portion of the SE component (see the map of [O{\sc iii}]$\lambda$4363 in Fig.~\ref{line_maps}).  The left panel of Fig.~\ref{te} displays the distribution of the
[O{\sc iii}] electron temperature which shows values going from near
15,000 K to $\gtrsim$ 22,000 K; here it is the first time that $T_{\rm
  e}$[O{\sc iii}] values $\gtrsim$ 22,000 K are derived for IZw18.
Such high values of $T_{\rm e}$[O{\sc iii}] are rarely found in nearby
HII galaxies/HII regions \citep[e.g.][]{EPMC09}.  

In active SF regions, stellar winds from massive stars and/or
supernovae remnants will likely produce shock-waves. As for IZw18, the
complex geometry of its ionized gas argue in favor of a contribution
due to shocks \citep[e.g.][]{HT95,cannon02}. Thus a
shock-contamination on emission lines cannot be ruled out. One of the
effects of shock-waves is to increase the [O{\sc
iii}]$\lambda$4363/[O{\sc iii}]$\lambda$5007 ratio, and therefore the
$T_{\rm e}$[O{\sc iii}] \citep[e.g.,][]{p91}.  As $T_{\rm
e}$[O{\sc iii}] $\gtrsim$ 20,000 K is more typical of shock-heated
nebulae than of SF regions \citep[e.g.,][]{b81,K11}, inititaly one could guess that our highest
values of $T_{\rm e}$[O{\sc iii}] are, in part, the result of shock
excitation. However, observational arguments reason against a
significant shock component to our data. We find no evidence for
[S{\sc ii}]/H$\alpha$ and/or [O{\sc
i}]/H$\alpha$ enhancement (a usual sign of
shock-excitation; e.g., \citealt{s85,d96}) associated with the higher-$T_{\rm e}$[O{\sc
iii}] spaxels (see Figs.~\ref{line_ratio_maps} and \ref{te});
actually, most of the [O{\sc i}] emission is
concentrated on the SE knot where we found relatively lower values of $T_{\rm
e}$[O{\sc iii}] (see
Fig.~\ref{te}). Besides, the BPT diagrams in Fig.~\ref{bpt}  indicate that shocks do not
play an important role in the gas excitation (see Section~\ref{sris}). So the enhanced [O{\sc iii}] temperatures
derived are expected to be associated primarily with photoionization from hot massive stars, and any possible contribution to the $T_{\rm e}$[O{\sc iii}] errors due to shocks should be negligible.

It is well known that in the process of flux measurement of very faint
emission lines, like [O{\sc iii}]$\lambda$4363, a bias (e.g. continuum
fitting) is expected to overestimate line
intensities \citep[e.g.,][]{RP94,K04}. To check if such a bias is
affecting our [O{\sc
iii}]$\lambda$4363 flux, we plot the relation between $T_{\rm
e}$[O{\sc iii}] and the relative error measured for the [O{\sc
iii}]$\lambda$4363 line in the right panel of Fig.~\ref{te}, which
shows no systematic effect. This indicates that the highest values of
$T_{\rm e}$[O{\sc iii}] that we compute are real and not an
effect of an overestimation during the measurement of the [O{\sc iii}]$\lambda$4363
flux. Note that we also derive lower values of $T_{\rm e}$[O{\sc
iii}] $\sim$ 16,000 - 18,000 K outside the two central knots,
consistent with results reported in previous work
\citep[e.g.,][]{jvm98}. All this supports the robustness of our
measurements of the  [O{\sc iii}]$\lambda$4363 line used to derive $T_{\rm e}$[O{\sc iii}].

For the low-excitation zone, no auroral line (e.g., [O{\sc
ii}]$\lambda$7320,7330; [N{\sc ii}]$\lambda$5755) could be measured in
any spaxel. In our case, the values of $T_{\rm e}$[O{\sc ii}] were
calculated from the empirical relation between [O{\sc ii}] and [O{\sc
iii}] electron temperatures given by \cite{PI06}. This relation has
been successfully used to calculate $T_{\rm e}$[O{\sc ii}] in other
HII galaxies \citep[e.g.,][]{K08,C09mrk1418,C10}.

The oxygen ionic abundance ratios, O$^{+}$/H$^{+}$ and
O$^{2+}$/H$^{+}$, were derived from the [O{\sc ii}]$\lambda$3727 and
\mbox{[O{\sc iii}]$\lambda\lambda$ 4959,5007} lines, respectively
using the corresponding electron temperatures.  A small fraction of
the unseen O$^{3+}$ ion is expected to be present in HII regions that
show high-ionization emission lines like He{\sc ii}$\lambda$4686 in
their spectra. According to the photoionization models from
\cite{I06}, the O$^{3+}$/O ratio is $>$ 1$\%$ only in the
highest-excitation HII regions for which O$^{+}$/(O$^{+}$ + O$^{2+}$)
$<$ 10$\%$. We have checked that for all He{\sc
  ii}$\lambda$4686-emitting spaxels\footnote{The map of He{\sc
    ii}$\lambda$4686 is shown in Fig. 2 from \cite{K15}},
O$^{+}$/(O$^{+}$ + O$^{2+}$) is higher than 10$\%$, so the total
oxygen abundance is assumed to be: \mbox{O/H = O$^{+}$/H$^{+}$ +
  O$^{2+}$/H$^{+}$}.  The map and histogram of the derived oxygen
abundance are displayed in Fig.~\ref{oh}. The calculated values of
12+log(O/H) vary approximately between 6.90 and 7.35, with more than 60$\%$ of the spaxels showing oxygen
abundances in the range of $\sim$ 7.0-7.2. 

As mentioned above, we
see differences of 
up to $\sim$ 10,000 K among our $T_{\rm e}$[O{\sc iii}] measurements (see Fig.~\ref{te}). At first, one
can think that such change in $T_{\rm e}$[O{\sc iii}]  might be mainly
the result of differences in the metallicity across IZw18. However, although some
degree of O/H variations are observed when considering
individual spaxels (see Fig.~\ref{oh}), we find that such variations are not statistically
significant, and that the ionized gas of IZw18 is chemically
homogeneous over spatial scales of hundreds of parsecs (see
Section~\ref{stat} for details). This points out that the observed
difference in $T_{\rm e}$[O{\sc iii}]  should be mostly due to changes in the
ionizing radiation field. Actually, close inspection of Fig.~\ref{te} indicates that most of 
the largest values of $T_{\rm e}$[O{\sc iii}] are located in the
 northwestern region of IZw18, where we find the gas excitation and
 the ionization parameter (as traced by [O{\sc iii}]/[O{\sc ii}]) to
 be higher too (see Section~\ref{sris}). It is worth mentioning that
 the  [O{\sc iii}]/[O{\sc ii}] ratio also depends on the effective
 temperature of the ionizing source(s) so that [O{\sc iii}]/[O{\sc
   ii}] is larger for higher effective
 temperatures \citep[e.g.,][]{jvm88,EPMD05}. Also, Fig.~\ref{o3o2_ro3} shows that there is a general trend
 for the line-ratio indicator of the $T_{\rm e}$[O{\sc iii}] to go up when the 
 [O{\sc iii}]/[O{\sc ii}] ratio increases. The fact that the youngest ionizing stars
 of IZw18 are mostly concentrated in the NW portion
 \citep[e.g.,][]{CR11} supports these results.  Additionaly, \citet[][]{K15}
report the presence of an extended emission in the high-excitation
He{\sc ii}$\lambda$4686 line which is spatially associated with the
NW cluster. These authors claim that peculiar very hot, 
(nearly) metal-free ionizing stars are
needed to account for the total He{\sc ii}-ionization budget in IZw18.
Moreover, we note here that more than 70$\%$  of the spaxels
with $T_{\rm e}$[O{\sc iii}] $\gtrsim$ 22,000 K are HeII-emitting
spaxels too. These facts reinforce the existence of a harder ionizing
radiation field at the location of the NW SF knot and thereabouts.

\section{Analysis of the spatial variation of physical conditions and chemical abundances}\label{stat}

Here, we make use of a statistical analysis to study the spatial
variation of properties of the ionized gas across our FOV. We assume
that a certain physical-chemical property is homogeneous over 
IZw18  if two conditions are satisfied: for the corresponding
dataset (i) the null hypothesis (i.e. the data come from a normally
distributed population) of the Lilliefors test \citep{li67} cannot be
rejected at the 10$\%$ significance level, and (ii) the observed
variations of the data distribution around the single mean value can
be explained by random errors; i.e. the corresponding Gaussian sigma
($\sigma_{Gaussian}$) should be lower or of the order of the typical
uncertainty of the property considered; we take as typical uncertainty
the square root of the weighted sample variance
($\sigma_{weighted}$). Previous work has successfully applied this
method to the analysis of abundances and physical conditions in HII
galaxies \citep[see][]{P11,K13,EPM13}. Table~\ref{fits} displays the
results from our statistical analysis for several relevant ISM properties.

From Table~\ref{fits}, we can see that the distribution of the
measured values for $T_{\rm e}$[O{\sc iii}], O/H, [O{\sc iii}]/[O{\sc ii}]
and R$_{23}$, from individual spaxels, can be represented by a Gaussian fit in agreement with
the Lilliefors test (i.e., the corresponding significance levels for
such distributions are $>$ 10$\%$).  Despite of that, in the case of
the $T_{\rm e}$[O{\sc iii}] and of the [O{\sc iii}]/[O{\sc ii}] ratio, we find
that $\sigma_{Gaussian}$ $>$ $\sigma_{weighted}$. This points out that
the $T_{\rm e}$[O{\sc iii}] and [O{\sc iii}]/[O{\sc ii}] values are not
homogeneously distributed across IZw18, and that, at first
order, random variables alone could not explain the
distribution observed of these two ISM properties.  This result gives
support to the existence of an electron temperature gradient in
IZw18 as suggested by \citet[][]{IT99} from their long-slit
spectroscopic analysis of the NW and SE components \citep[see
also][]{jvm98}.

Regarding the O/H spatial distribution, the two conditions for a given
property to be considered homogeneous, as mentioned above, are
accomplished: the derived values of O/H are fitted by a normal
distribution according to the Lilliefors test, and the corresponding
$\sigma_{Gaussian}$ is of the order of $\sigma_{weighted}$. Our
results, therefore, show that the ionized gas-phase O/H remains mostly
uniform over spatial scales of hundreds of parsecs. This confirms that
there is no significant abundance gradient nor measurable
discontinuity in IZw18, as it has been suggested in previous work
\citep[][]{jvm98,L00}. Here, we assume that the representative
metallicity of IZw18 is 12+log(O/H) = 7.11 $\pm$ 0.01 ($\sim$ 1/40 of
the solar metallicity) which represents the derived error-weighted
mean value of O/H and its corresponding statistical error from all the
individual spaxel O/H abundances obtained using the direct method with
electron temperature measurement (for this O/H distribution, the
square root of the weighted sample variance associated is 0.12 dex;
see Table~\ref{fits}).  The observed absence of significant metallicity
difference among the areas near the two main SF knots and the gas
located farther out impose strong constrains for the mechanisms that
triggered star-formation and also for the chemical evolution history of
IZw18. The HII regions in IZw18 should have evolved  along a similar enrichment
scenario over the whole scale of the galaxy  in order to produce the observed chemically homogeneous ISM
\citep[e.g.,][]{K95,RK95,jvm98}.

Following the same statistical analysis, the behaviour of the R$_{23}$
distribution is observed to be similar to that of the oxygen
abundance, i.e., the R$_{23}$ parameter seems to be very uniform
across IZw18, thus being insensitive to possible effects from changes
in the hardness of the ionizing radiation and the ionization parameter (as traced by [O{\sc iii}]/[O{\sc
  ii}]). This result indicates that R$_{23}$ can be considered as a
good metallicity indicator in metal-poor SF galaxies \citep[see
e.g.,][]{pagel79,jvm95,P00,J15}. Indeed, the constant metallicity
derived over the entire scale of IZw18, despite the non-homogeneous
distribution in the degree of ionization observed, clearly shows its
independence from the ionization parameter (see also
Section~\ref{sris}).

\begin{table*}
\centering
\caption{Results from the statistical analysis for the distribution
  of the physical-chemical conditions in the ISM across the PMAS FOV of IZw18.}
\label{fits}
\begin{minipage}{9.0cm}
\begin{tabular}{lccccc}
\hline\hline \\
& \multicolumn{5}{c}{Statistical properties} \\
 & $\mu_{weighted}$$^{a}$ & $\sigma_{weighted}$$^{b}$ & $\mu_{Gaussian}$$^{c}$ & $\sigma_{Gaussian}$$^{d}$ & Sign.(\%)$^{e}$  \\
& & & & &\\ \hline
$T_{\rm e}$([O{\sc iii}]) (K) & 21300 & 2400  & 22400  & 4180  & 36  \\ 
12+log(O/H)         & 7.11  & 0.12  & 7.06  & 0.14 & 34 \\ 
log([O{\sc iii}]/[O{\sc ii}])  & 0.55  & 0.24  & 0.60  & 0.33 & 25  \\ 
log($R_{23}$)       & 0.47 & 0.05  & 0.47 & 0.06 & 50 \\ \hline
\end{tabular}
\end{minipage}
 \begin{flushleft}
(a) error-weighted mean;(b) square root of the weighted sample variance associated with
the corresponding weighted mean; (c) mean of the Gaussian
distribution; (d) standard deviation of the Gaussian distribution; (e)
significance level of the null hypothesis in the Lilliefors test (see the text for details). 
\end{flushleft}
\end{table*}

\section{Properties of selected regions of IZw18 from integrated spectra}\label{int}

We also take advantage of our IFU data to simulate the 1D spectra of
selected galaxy regions.  The representative spectra of the NW and SE
knots, and of the ``plume'' region were constructed by adding the flux
in the spaxels within each of the corresponding areas whose boundaries
are displayed on the H$\alpha$ map in Fig.~\ref{line_maps}. For both
the NW and SE knots, the area of their aperture extraction is of 3
arcsec $\times$ 3 arcsec; the ``plume''-integrated region corresponds
to an area of $\sim$ 12 arcsec$^{2}$. In addition, we have created a
1D spectrum representative of the gaseous ionized ``halo'' of IZw18.
To do so we have integrated the flux in all the spaxels for which the
H$\alpha$ flux measurements present a relative error $\gtrsim$ 20$\%$;
thus exclusively the emission of the fainter ionized gas (``halo'')
has been integrated in the halo spectrum.  We also obtained, for the
first time, the IZw18 integrated spectrum by summing the emission from
each spaxel within an area of $\sim$ 168 arcsec$^{2}$ ($\sim$ 1.3
kpc$^{2}$ at the distance of 18.2 Mpc), enclosing basically all the
nebular emission across our FOV. Considering IZw18 as an excellent local analog of primeval systems,
the analysis of its integrated properties may be important for the study of
intermediate/high redshift SF galaxies for which only their integrated
characteristics are known due to their
distance \citep[e.g.,][]{kew13,nakajima13}.

\begin{figure*} 
\includegraphics[width=8.5cm,clip]{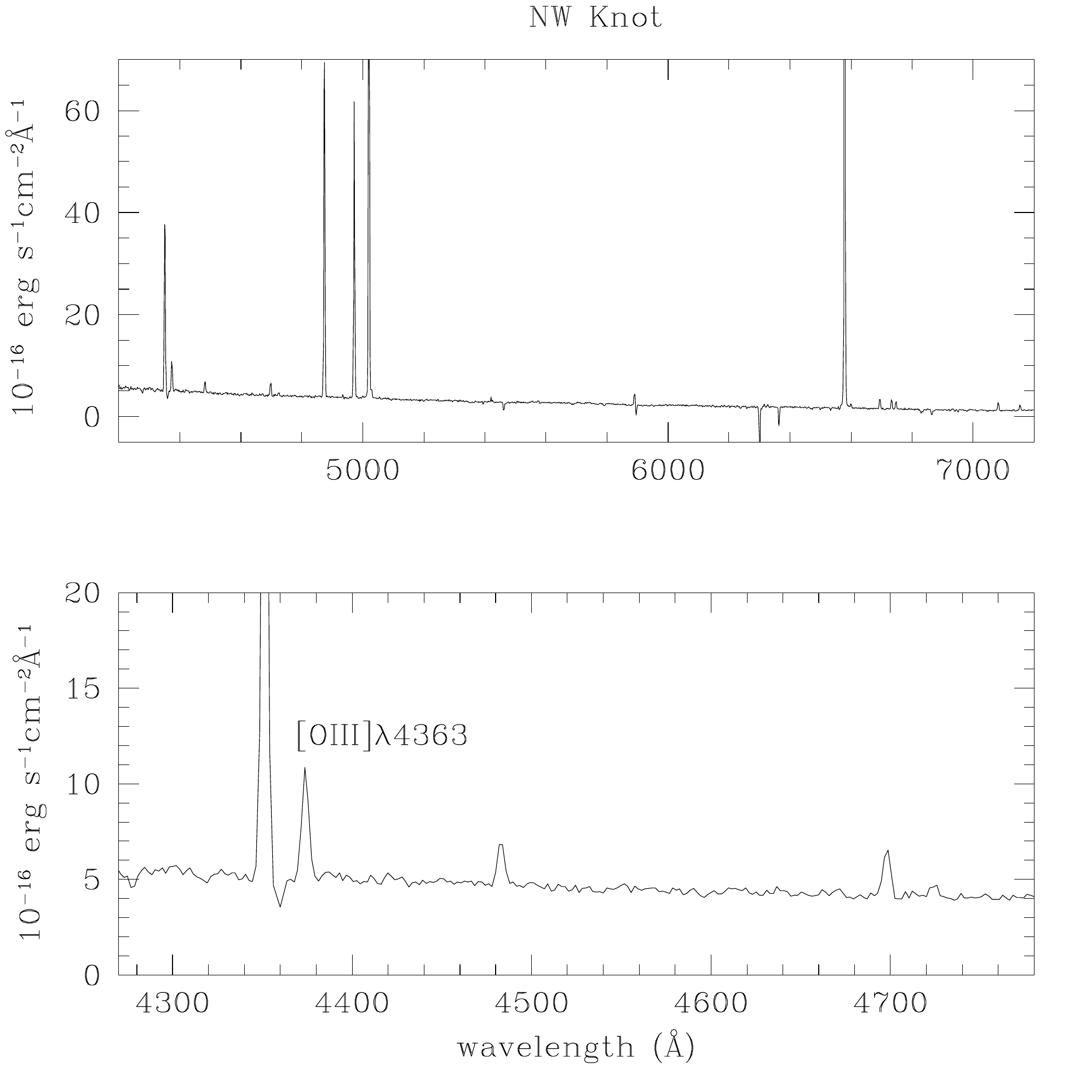} 
\includegraphics[width=8.5cm,clip]{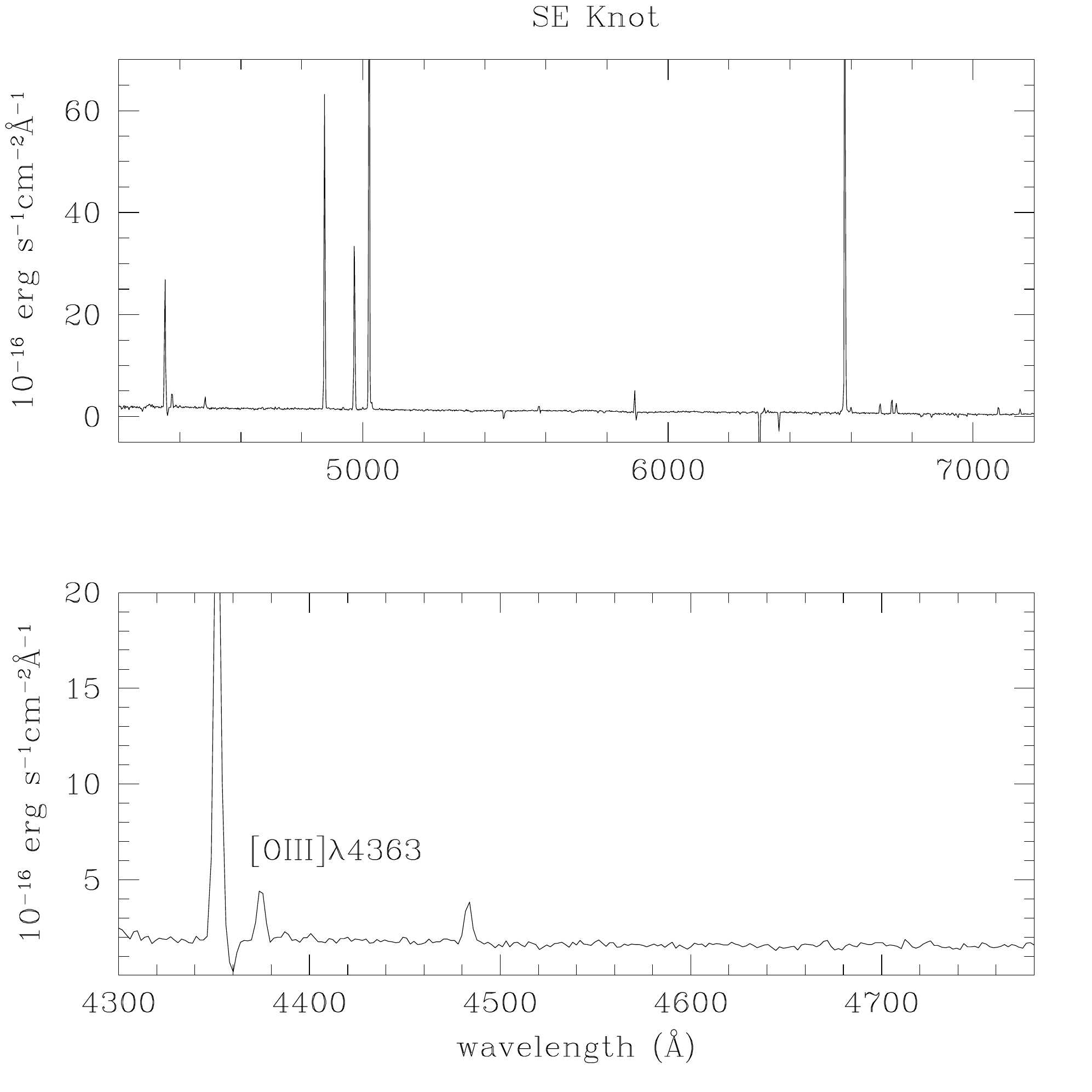} \\
\includegraphics[width=8.5cm,clip]{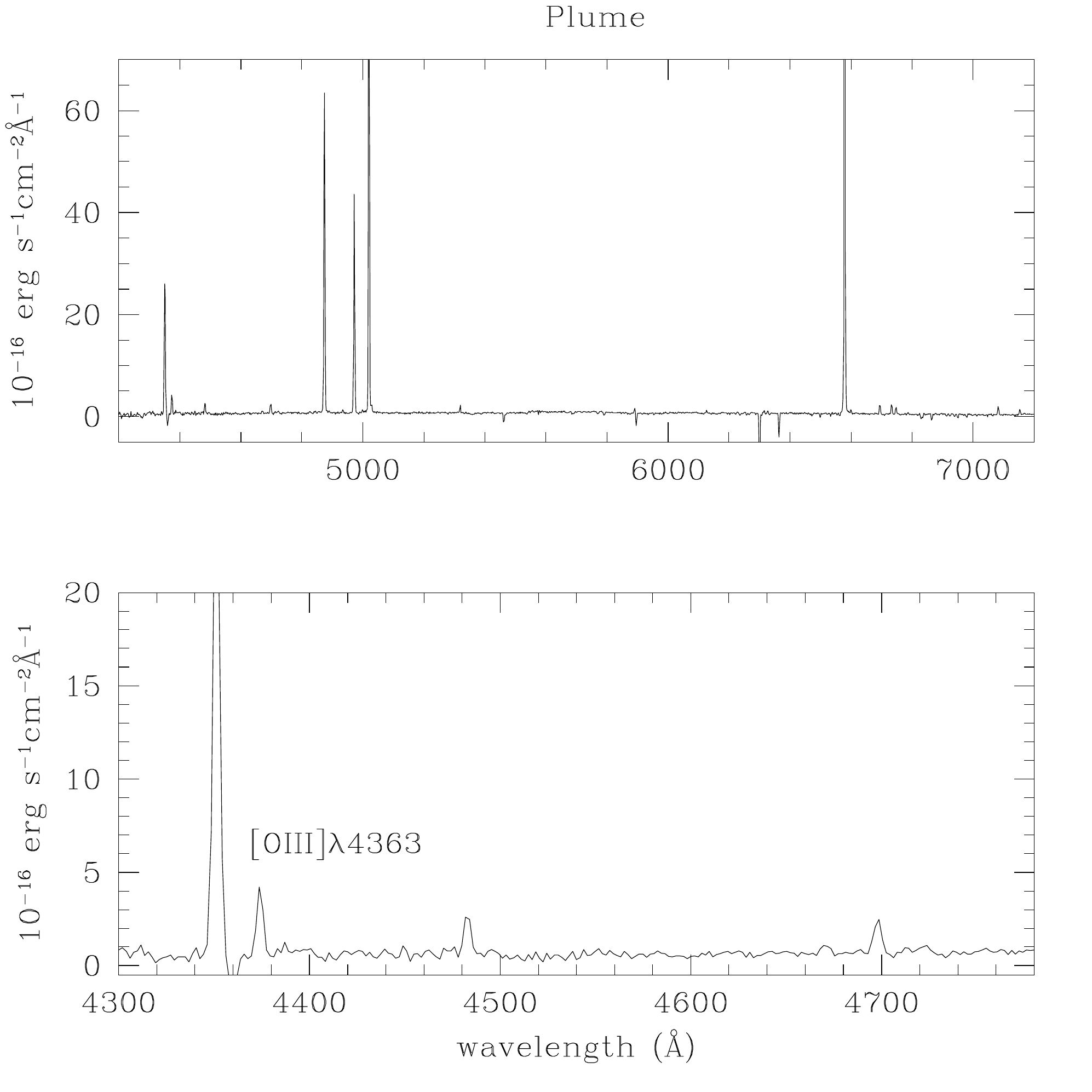} 
\includegraphics[width=8.5cm,clip]{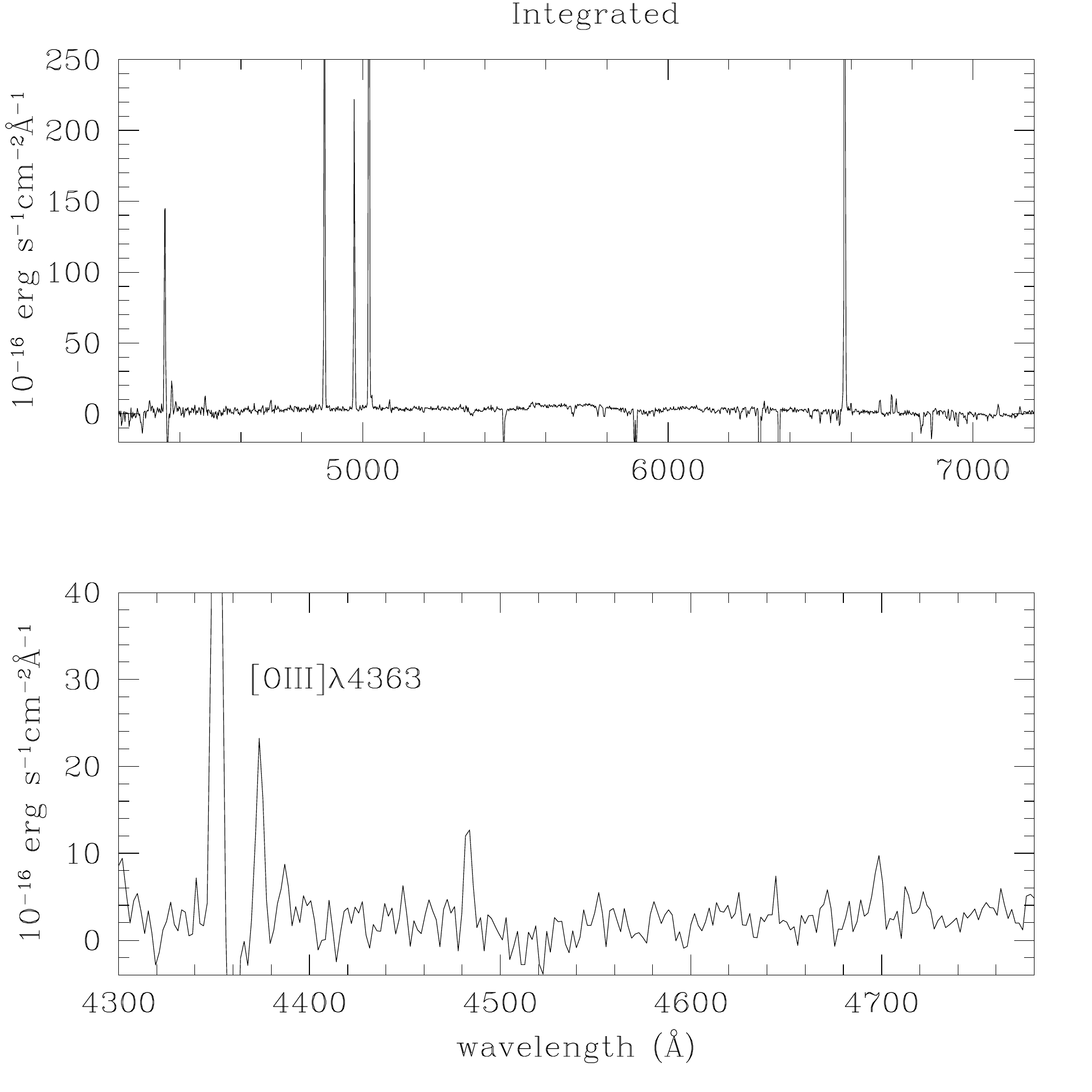} \\
\includegraphics[width=8.5cm,clip]{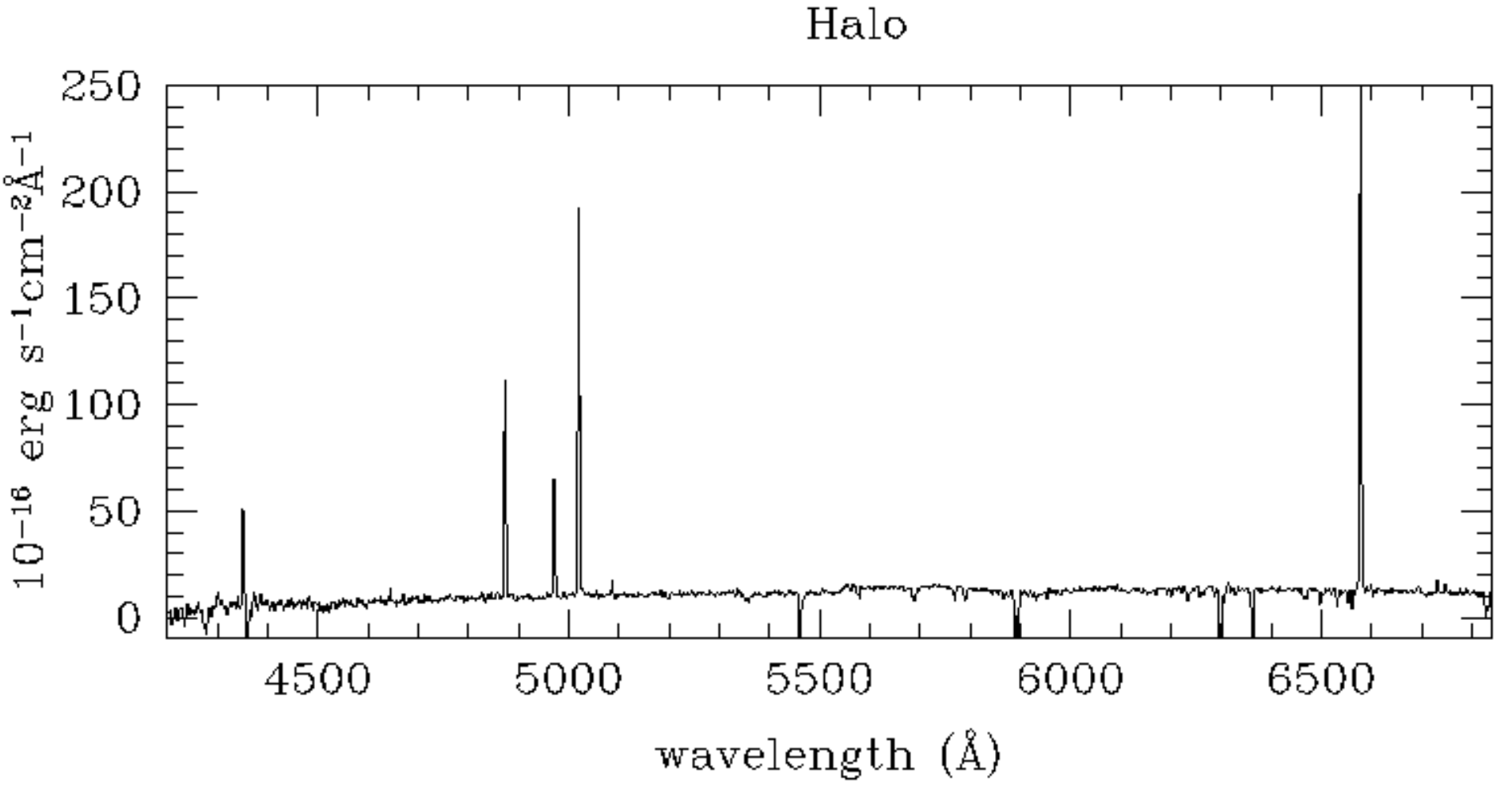}
\caption{Flux-calibrated 1D spectra of the selected regions of IZw18 (see the
text for details). A zoom of the wavelength range $\sim$
4300-4700 \AA~showing the temperature sensitive emission line
[OIII]$\lambda$4363 for the NW knot, SE knot, ``plume'' region and
IZw18 integrated are also displayed. The spectra are in units of 10$^{-16}$ erg s$^{-1}$ cm$^{-2}$ \AA$^{-1}$.}
\label{1dspectra} 
\end{figure*}

Fig.~\ref{1dspectra} presents the 1D spectra for the aforementioned
regions of IZw18. The emission-line fluxes and c(H$\beta$)
corresponding to each summed spectra were measured following the same
procedure as in Section~\ref{flux_measurements}. For the integrated
spectra, we derived the physical conditions and oxygen abundances as
described in Section~\ref{2dgas} for individual spaxels. The nitrogen
ionic abundance ratio, N$^{+}$/H$^{+}$, was calculated from the
PYNEB-based expression given by \cite{EPM14}, using the [N{\sc ii}]$\lambda$6584
emission line and assuming T$_{e}$[N{\sc ii}] $\sim$
T$_{e}$[O{\sc ii}]; the N/O abundance ratio was computed under the
assumption that N/O = N$^{+}$/O$^{+}$, based on the similarity of the
ionization potentials of the ions involved. Reddening-corrected line intensities, normalized to
H$\beta$, along with the physical-chemical properties derived from the
integrated spectra are shown in Table~\ref{table_regions}.

The measurements of the integrated line-ratios [O{\sc
  iii}]$\lambda$5007/H$\beta$, [N{\sc ii}]$\lambda$6584/H$\alpha$,
[S{\sc ii}]$\lambda$6717,6731/H$\alpha$ and [O{\sc
  i}]$\lambda$6300/H$\alpha$  place all the selected
regions on the SF zone of the BPT diagrams (see Fig.~\ref{bpt}).
The comparison among the values for the integrated [O{\sc
iii}]/[O{\sc ii}] ratios shows that the NW knot and ``plume'' spectra
present  [O{\sc iii}]/[O{\sc ii}] higher than the ratios derived for the other
regions (see Table~\ref{table_regions}). This result is consistent with the presence of a harder
ionizing radiation field near the northwestern area of IZw18, as
discussed in Section~\ref{2dgas}. 

The $T_{\rm e}$[O{\sc iii}] and O/H derived for the NW and SE
components are consistent, within the errors, with those obtained
by \citet[][]{IT99} through long-slit spectroscopy along the NW-SE
knots. In comparison with the values of 12 + log(O/H)
from \citet[][]{SK93} and \citet[][]{jvm98}, our oxygen abundances ([O{\sc iii}]
electron temperatures) are
somewhat lower (higher) than theirs. The
discrepancy between our values and those previously obtained based on
long-slit spectroscopy may be due to their spectra covering partially
different regions due to slit position and orientation. Table~\ref{tablesummary} summarizes, for both the NW and SE knots, the $T_{\rm e}$[O{\sc iii}] and O/H measurements derived here and those formerly reported in the references mentioned above.

By comparing the different electron temperature values from
Table~\ref{table_regions}, we find that the NW knot shows a relatively
higher temperature, in agreement with the observed spatial distribution
of $T_{\rm e}$[O{\sc iii}] (see Section~\ref{2dgas}). Regarding the oxygen abundances, the SE knot, ``plume'' and
IZw18-integrated spectra give equivalent values 
considering the corresponding error bars; the NW knot presents a
barely lower O/H. Remarkably, however, we note that 12+log(O/H)
derived from all selected region spectra agree with each other within $\sim$ 0.10 dex
which is of the order of $\sigma_{weighted}$ (see Tables~\ref{fits}
and \ref{table_regions}). This means that the integrated-spectrum
metallicty [12+log (O/H)=7.10 $\pm$ 0.03] is consistent with the O/H
from the other selected regions (which cover smaller areas) and with
the representative metallicity of IZw18 based on spaxel-by-spaxel
measurements as defined previously in Section~\ref{stat}. Thus, our IFS
study shows that O/H of IZw18 does not depend on the aperture size
used and that the IZw18 integrated spectrum can reliably represent the
metallicity of its ionized ISM. 

With respect to the nitrogen
abundance, we find that all the selected regions of IZw18 present similar N/O
ratios within the uncertainties (see
Table~\ref{table_regions}). The N/O values computed here match the
characteristic N/O [-log(N/O) $\sim$ 1.5-1.6] observed for low-metallicity
systems which form the well-known N/O plateau \citep[e.g.,][]{IT99,molla06,P11}.  Additionaly, we checked
that the N/O ratios for the NW and SE knots agree with those
reported in previous studies \citep[e.g.,][]{SK93,jvm98,IT99}

\begin{table*}
\caption{De-reddened emission line-fluxes relative to 1000$\cdot$I(H$\beta$) 
and physical properties from different selected regions} 
\label{table_regions}
 \centering 
\begin{minipage}{15.0cm}
\centering
\begin{tabular}{lccccc}
\hline\hline 
Wavelength (\AA) & NW Knot  & SE Knot  & ``Plume'' & ``Halo'' & Integrated     \\  \hline
3727 [O~II]    & 374 $\pm$ 16 & 678 $\pm$ 21 & 372 $\pm$ 25 & 714 $\pm$ 40 & 581 $\pm$ 50  \\ 
3868 [Ne~III]  & 179 $\pm$ 6  & 145 $\pm$ 5  & 128 $\pm$ 8    & --- & 101 $\pm$ 17 \\
4340 H$\gamma$ & 526 $\pm$ 8  & 488 $\pm$4   & 467 $\pm$ 4 & 476 $\pm$ 18 & 490 $\pm$ 7 \\
4363 [O~III]  &  87 $\pm$ 4   & 55 $\pm$ 3   & 56  $\pm$ 2 & --- & 63 $\pm$ 6 \\ 
4471 He~I      & 34 $\pm$ 1   & 38 $\pm$ 2   & 36  $\pm$ 3 & --- & 38 $\pm$ 4\\
4686 He~II    &  40 $\pm$ 3   & ---          & 34 $\pm$ 3  & --- & 23 $\pm$ 4 \\
4714 [Ar~IV]  &  9 $\pm$ 2    & ---          & --- & --- & --- \\
4861 H$\beta$ & 1000 $\pm$ 1  & 1000 $\pm$ 4 & 1000 $\pm$ 6 & 1000 $\pm$ 15  & 1000 $\pm$ 7 \\
4959 [O~III]  & 741 $\pm$ 13  & 605 $\pm$ 6  & 629  $\pm$ 7 & 555 $\pm$ 10 & 625 $\pm$ 8 \\
5007 [O~III]  & 2093 $\pm$ 10 & 1697 $\pm$ 19& 1787 $\pm$ 16 & 1650 $\pm$ 29 & 1800$\pm$ 21 \\
6300 [O~I]    &  7 $\pm$ 1    & 15 $\pm$ 1   & 11 $\pm$ 2 & --- & ---\\
6312 [S~III]  &  6 $\pm$ 1    & 6 $\pm$ 1    & 7 $\pm$ 1 & --- &  ---\\
6563 H$\alpha$ & 2750 $\pm$ 7 & 2750 $\pm$ 14 & 2750 $\pm$ 17 & 2576 $\pm$ 37  & 2750 $\pm$ 20 \\
6584 [N~II]    & 8.39 $\pm$ 0.60   & 17 $\pm$ 1    & 8.30  $\pm$ 0.40  & --- & 13 $\pm$ 2 \\
6678 HeI       & 27 $\pm$ 1   & 30 $\pm$ 2    & 28  $\pm$ 2 & --- & 29 $\pm$ 2 \\
6717 [S~II]    & 26 $\pm$ 1   & 46 $\pm$ 2    & 29  $\pm$ 2 & 52 $\pm$ 4 & 41 $\pm$ 2  \\ 
6731 [S~II]    & 20 $\pm$ 1   & 34 $\pm$ 2    & 20  $\pm$  2 & 27  $\pm$ 4 & 28 $\pm$ 3\\  
7065 He~I      & 24 $\pm$ 1   & 25 $\pm$ 1    & 25 $\pm$ 1  & --- & 27 $\pm$ 2\\
7135 [Ar~III]  & 17 $\pm$ 1   & 16 $\pm$ 1    & 16 $\pm$ 2 &  --- & 16 $\pm$ 3 \\ \hline
c(H$\beta$)    & 0.13  & 0.13   & 0.09   & 0.00  & 0.04 \\
-EW(H$\beta$) (\AA)  & 76   & 150    & 320   & 23 & 350 \\
F(H$\beta$) (erg s$^{-1}$ cm$^{-2}$) & 3.95$\times10^{-14}$ & 3.23$\times10^{-14}$ & 3.16$\times10^{-14}$ & 4.50$\times10^{-14}$ & 1.59$\times10^{-13}$ \\
F(H$\alpha$) (erg s$^{-1}$ cm$^{-2}$) & 1.09$\times10^{-13}$ & 8.87$\times10^{-14}$ &8.70$\times10^{-14}$ & 1.16$\times10^{-13}$& 4.36$\times10^{-13}$\\  \hline
log ($R_{23}$) & 0.50  & 0.47  & 0.44 & 0.46 & 0.47 \\
log ([O~III]/[O~II]) & 0.87  & 0.52  & 0.81  & 0.49 & 0.61 \\
log ([O~I]6300/H$\alpha$) & -2.59  & -2.27  & -2.41  & --- & --- \\
log ([N~II]6584/H$\alpha$) & -2.52  & -2.21  & -2.52 & --- & -2.31 \\
log ([S~II]6717+6731/H$\alpha$) & -1.78 & -1.53  & -1.74 & -1.52 & -1.60 \\
log ([O~III]5007/H$\beta$) & 0.32  & 0.23  &  0.26 & 0.22 & 0.26 \\
$n_{\rm e}$([S~II])(cm$^{-3}$)  & 110   & $<$ 100  &$<$ 100 & $<$ 100 &$<$ 100 \\
$T_{\rm e}$([O~III]) (K) & 23000 $\pm$ 700  & 19600 $\pm$ 600  & 19200 $\pm$ 500 & --- & 20600 $\pm$ 1300 \\
$T_{\rm e}$([O~II])$^{a}$ (K) & 19200 $\pm$ 500 & 16700 $\pm$ 500 & 16400 $\pm$ 400 & --- & 17400 $\pm$ 900 \\
12+log($O^{++}/H^{+}$)  & 6.96 $\pm$ 0.02  & 7.01 $\pm$ 0.02  &  7.05 $\pm$ 0.02 & --- & 6.99 $\pm$ 0.04  \\ 
12+log($O^{+}/H^{+}$) & 6.17 $\pm$ 0.03 &  6.59 $\pm$ 0.03  & 6.35 $\pm$ 0.04  & --- & 6.47 $\pm$ 0.06  \\
12+log(O/H)$_{T_{\rm e}}$$^{b}$ & 7.03 $\pm$ 0.02  &7.15  $\pm$ 0.02  & 7.13 $\pm$ 0.02 & --- & 7.10 $\pm$ 0.03 \\ 
12+log($N^{+}/H^{+}$)  & 4.65 $\pm$ 0.03  & 5.06 $\pm$ 0.03  & 4.76 $\pm$ 0.03 & --- & 4.92 $\pm$ 0.06 \\
log(N/O)$_{T_{\rm e}}$$^{c}$    & -1.52  $\pm$ 0.05 & -1.53 $\pm$ 0.04 & -1.58  $\pm$ 0.04 & --- & -1.55  $\pm$ 0.09 \\
\hline
\end{tabular}
\end{minipage}
\begin{flushleft}
(a) $T_{\rm e}$([O{\sc ii}]) = 0.72 $\times$ $T_{\rm e}$ ([O{\sc
  iii}])+ 0.26 \citep{PI06}\\
(b) O/H = (O$^{+}$/H$^{+}$ + O$^{2+}$/H$^{+}$)\\
(c) N/O = $N^{+}$/O$^{+}$
\end{flushleft}
\end{table*}

 \begin{table*} 
\caption{Summary of [O{\sc iii}] electron temperatures and oxygen abundances for the NW and SE components} 
\label{tablesummary} 
\centering 
\begin{minipage}{14.9cm} 
\centering 
\begin{tabular}{lccc} 
\hline\hline  
Property & NW knot & SE Knot & Reference$^{*}$
\\ \hline 
\multirow{4}{*}{$T_{\rm e}$([O~III]) (K)} & 19600 $\pm$ 900 & 17200 $\pm$ 1200  & 1 \\  
                                        & 20300 $\pm$ 700 & 17700 $\pm$ 600 & 2 \\
                                        & 21500 $\pm$ 1400 & 19500 $\pm$ 800 & 3\\
                                        & 23000 $\pm$ 700  & 19600 $\pm$ 600  &  4 \\  \hline

\multirow{4}{*}{12+log(O/H)$_{T_{\rm e}}$} & 7.17 $\pm$ 0.04 & 7.26 $\pm$ 0.05 & 1  \\  
                                        & 7.16 $\pm$ 0.05  & 7.32 $\pm$ 0.08 & 2   \\
                                        & 7.07 $\pm$ 0.05 & 7.17  $\pm$ 0.06 & 3  \\ 
                                        & 7.03 $\pm$ 0.02 & 7.15  $\pm$ 0.02 & 4   \\ \hline 
                                         
\hline 
\end{tabular} 
\end{minipage}
\begin{flushleft}
* References: (1) \cite{SK93}; (2) \cite{jvm98}; (3) \cite{IT99}; (4) This work

\end{flushleft}
\end{table*} 
 
\section{Summary and conclusions}

We have analysed PMAS-IFU integral field spectroscopy of IZw18, an
extremely low metallicity galaxy, which is our best local laboratory
for probing the conditions dominating in distant metal-poor
starbursts.  These data map the entire spatial extent of the IZw18
main body plus an important region of the extended ionized gas,
providing us with a new 2D view of the ionized ISM in IZw18. Maps for the
spatial distribution of relevant emission lines and of physical-chemical
properties for the ionized gas have been created and analysed. We
believe that our observations provide a useful test-bench for realistic
photoionization models at the lowest metallicity regime.

Our spaxel-by-spaxel analysis indicates that despite the observed large range of
values ($\sim$ 0.0 to 0.9 dex) for the log [O{\sc iii}]/[O{\sc ii}] 
ratio (widely used as an ionization parameter indicator), the
metallicity index R$_{23}$ remains substantially uniform; no
dependence 
between R$_{23}$ and the ionization parameter is seen in IZw18.  The
BPT diagrams, [N{\sc ii}]$\lambda$6584/H$\alpha$, [S{\sc
  ii}]$\lambda$6717,6731/H$\alpha$, and [O{\sc
  i}]$\lambda$6300/H$\alpha$, for the spatially resolved emission
lines, indicates that photoionization by massive stars is the dominant
excitation source for the gas within our FOV. We have detected a very
faint blue bump of WR stars towards the NW knot. Regions with higher
excitation and harder ionizing radiation
are preferentially located towards the NW zone of IZw18 where we find the
largest values of $T_{\rm e}$[O{\sc iii}] too. Also, the NW component
is spatially related to an extended He{\sc ii}$\lambda$4686-emitting
region that has been proposed to indicate the presence of peculiar very hot, ionizing stars which may be
(nearly) metal-free \citep[see][]{K15}. 

Our statistical analysis shows an important degree of non-homogeneity
for the $T_{\rm e}$[O{\sc iii}] distribution and that the scatter in
$T_{\rm e}$[O{\sc iii}] can be larger than that in O/H within the
observed [O{\sc iii}]$\lambda$4363-emitting region (42 arcsec$^{2}$
$\sim$ 0.3 kpc$^{2}$ at the distance of 18.2 Mpc). We find no
statistically significant variations in O/H across the 16'' $\times$
16'' PMAS-IFU aperture, indicating a global homogeneity of the
oxygen abundances in IZw18 over spatial scales of hundreds of
parsecs. The representative metallicity of IZw18 derived here, from individual
spaxel measurements, is 12 + log(O /H) =7.11 $\pm$ 0.01
(error-weighted mean value of O/H and its corresponding statistical
error). The prevalence of a substantial
degree of homogeneity in O/H over the IZw18 galaxy can constrain its
chemical history, suggesting an overall enrichment phase previous to
the current burst.

We took advantage of our IFU data to create 1D integrated spectra for
regions of interest in the galaxy. For the first time, we derive the
IZw18 integrated spectrum by summing the spaxels over the whole
FOV. Physical-chemical properties of the ionized gas were derived
from these selected region spectra. In the three BPT diagrams, all the
integrated regions fall within the area corrresponding to HII-like
ionization.  Putting together the spatially resolved measurements and
integrated ones of O/H, we find that the IZw18 integrated spectrum-O/H
concurs with the representative metallicity of IZw18 defined here, and
can be taken to describe the abundance of the ionized gas of IZw18 as a
whole. We also show that the derivation of O/H does not depend on the aperture size
used. This is a relevant result for studies of high-z SF metal-poor objects for
which only the integrated spectra are available.

\section*{Acknowledgements}
We are very grateful to our referee for providing constructive
comments and help in improving the manuscript. We wish to thank Martin
M. Roth as the PI of the PMAS IFU at the CAHA 3.5 m which allowed us
to carry out the first IFS study of the landmark galaxy IZw18. We also
thank the CAHA staff for their help during the observations. This work
has been partially funded by research projects AYA2010-21887-C04-01
and AYA2013-47742-C04-01 from the Spanish PNAYA, and PEX2011-FQM7058
from Junta de Andalucia. J.D.H.F. acknowledges support through the
FAPESP grant Project 2012/13381-0.

\bibliographystyle{mn2e}

\label{lastpage}
\end{document}